%
\input harvmac.tex

\def\mnu{$n_1$--$n_2$--$1$}
\def\tnu{\tilde n_1}
\def\tnd{\tilde n_2}
\def\La#1{\Lambda_{#1}}
\def\la#1{\lambda^{#1}}
\def\mpk{m_{\rm Pl}}
%
\def\npb#1#2#3{Nucl. Phys. {\bf B#1} (#2) #3}
\def\plb#1#2#3{Phys. Lett. {\bf #1B} (#2) #3}
\def\prd#1#2#3{Phys. Rev. {\bf D#1} (#2) #3}

\def\hth#1#2#3#4#5#6#7#8{hep-th#1#2#3#4#5#6#7#8}

\lref\NAD{
N. Seiberg, {\it Electric-Magnetic Duality in Supersymmetric
Non-Abelian Gauge Theories}, \npb{435}{1995}{129}, \hth/9411149.}

\lref\Seibone{
N. Seiberg,
{\it Exact Results on the Space of Vacua of
Four Dimensional supersymmetry Gauge Theories},
\prd{49}{1994}{6857}, \hth/9402044.}

\lref\emop{
R.G. Leigh and M.J. Strassler,
{\it Exactly Marginal Operators and Duality in
Four Dimensional N=1 Supersymmetric Gauge Theory},
\npb{447}{1995}{95}, \hth/9503121.}

\lref\ads{
I. Affleck, M. Dine and N. Seiberg,
{\it Dynamical Supersymmetry Breaking in Supersymmetric QCD},
\npb{241}{1984}{493};
{\it Dynamical Supersymmetry Breaking in Four Dimensions
and its Phenomenology}, \npb{256}{1985}{557}.}

\lref\dds{
A.C. Davis, M. Dine and N. Seiberg,
\plb{125}{1983}{487}.}

\lref\dnns{
M. Dine, A. E. Nelson, Y. Nir and Y. Shirman,
{\it New Tools for Low Energy Dynamical Supersymmetry Breaking},
\prd{53}{1996}{2658}, \hth/9507378.}

\lref\adsI{I. Affleck, M. Dine and N. Seiberg, {\it Dynamical Supersymmetry
Breaking in Chiral Gauge Theories}, \plb{137}{1984}{187}.}

\lref\meve{Y. Meurice and G. Veneziano, {\it SUSY Vacua Versus Chiral 
Fermions}, \plb{141}{1984}{69}.}

\lref\CERN{
D. Amati, K. Konishi, Y. Meurice, G.C. Rossi and G. Veneziano,
\phr{162}{1988}{169}, and references therein.}

\lref\intpoul{
K. Intriligator and P. Pouliot,
{\it Exact Superpotentials, Quantum Vacua and Duality in Supersymmetric
$Sp(N_c)$ Gauge Theories}, \plb{353}{1995}{471}, \hth/9505006.}

\lref\kinsso{K. Intriligator and N. Seiberg,
{\it Duality, Monopoles, Dyons, Confinement and Oblique Confinement in
Supersymmetric $SO(N_c)$ Gauge Theories},
\npb{444}{1995}{125}, \hth/9503179.}

\lref\kinsrev{K. Intriligator and N. Seiberg,
{\it Lectures on Supersymmetric Gauge Theories and Electric-Magnetic
 Duality,} Nucl.~Phys.~Proc.~Suppl. {\bf 45BC} (1996) 1,
 \hth/9509066.}

\lref\kutschw{D. Kutasov,
{\it A Comment on Duality in $N=1$ Supersymmetric Non-Abelian Gauge Theories},
\plb{351}{1995}{230}, \hth/9503086;
D. Kutasov and A. Schwimmer,
{\it On Duality in Supersymmetric Yang-Mills Theory},
\plb{354}{1995}{315}, \hth/9505004.}

\lref\dkasns{D. Kutasov, A. Schwimmer and N. Seiberg,
{\it Chiral Rings, Singularity Theory and Electric-Magnetic Duality},
\npb{459}{1996}{455}, \hth/9510222.}

\lref\ilslist{K. Intriligator, {\it New RG Fixed Points and
Duality in Supersymmetric $Sp(N_c)$ and $SO(N_c)$ Gauge
Theories}, \npb{448}{1995}{187}, \hth/9505051;
R.G. Leigh and M.J. Strassler, {\it Duality of
$Sp(2N_c)$ and $SO(N_c)$ Supersymmetric Gauge Theories with
Adjoint Matter}, hep-th/9505088, \plb{356}{1995}{492}.}

\lref\ilsmods{
K. Intriligator, R.G. Leigh and M.J. Strassler,
{\it New Examples of Duality in Chiral and Non-Chiral Supersymmetric
Gauge Theories}, \npb{456}{1995}{567}, \hth/9506148.}

\lref\berkooz{M. Berkooz,
{\it The Dual of Supersymmetric $SU(2k)$ with an Antisymmetric Tensor
and Composite Dualities},
\npb{452}{1995}{513}, \hth/9505067.}

\lref\mnmods{
C.~Cs\'{a}ki, L.~Randall and W.~Skiba,
{\it More Dynamical Supersymmetry Breaking}
\npb{479}{1996}{65}, \hth/9605108;
C.~Cs\'{a}ki, R.G.~Leigh, L.~Randall and W.~Skiba,
{\it Supersymmetry Breaking Through Confining and Dual Gauge Theory
Dynamics}, \plb{387}{1996}{791}, \hth/9607021.
}

\lref\iss{K. Intriligator,  N. Seiberg and S. Shenker 
{\it Proposal for a Simple Model of Dynamical supersymmetry Breaking},
\plb{342}{1995}{152}, \hth/9410203.}

\lref\aenns{A.E. Nelson and N. Seiberg,
{\it R Symmetry Breaking Versus Supersymmetry Breaking},
\npb{416}{1994}{46}, \hth/9309299. }

\lref\pts{E. Poppitz, Y. Shadmi and S.P. Trivedi,
{\it Supersymmetry Breaking and Duality in $SU(N)\times SU(N-M)$ Theories},
\plb{388}{1996}{561}, \hth/9606184;
{\it Duality and Exact Results in Product Group Theories},
\npb{480}{1996}{125}, \hth/9605113.}

\lref\ahem{ K. Intriligator and S. Thomas,
{\it Dynamical Supersymmetry Breaking on Quantum Moduli Spaces},
\npb{473}{1996}{121}, \hth/9603158.}

\lref\itdual{K. Intriligator and S. Thomas,
{\it Dual Descriptions of Supersymmetry Breaking},
\hth/9608046, SLAC-PUB-7143.}

\lref\shirman{Y. Shirman,
{\it Dynamical Supersymmetry Breaking versus Run-away behavior in
Supersymmetric Gauge Theories }, \plb{389}{1996}{287}, \hth/9608147.}

\lref\chau{}

%
%
\def\adj{\Sigma}\def\ad{\Sigma}\def\tadj{\tilde\adj}
\def\bF{\bar{F}}\def\fo{\bar{F_1}}\def\ft{\bar{F_2}}
\def\bP{\bar{P}}\def\bp{\bar{p}}
\def\brf{\bar{f}}
\def\l{\lambda_{1}}
\def\tl{\lambda_{2}}
\def\ttl{\lambda_{3}}
\def\gl{g_{1}}
\def\gtl{g_{2}}
\def\gttl{g_{3}}
\def\tr{{\rm tr}}
\def\xo#1{X_1^{#1}}
\def\xt#1{X_2^{#1}}
\def\d#1{\delta_{#1}}

\def\rk{\rm rank}

\def\gsim{{~\raise.15em\hbox{$>$}\kern-.85em
          \lower.35em\hbox{$\sim$}~}}
\def\lsim{{~\raise.15em\hbox{$<$}\kern-.85em
          \lower.35em\hbox{$\sim$}~}}

%
\font\zfont = cmss10 
\font\zfonteight = cmss8 
\def\ZZ{\hbox{\zfont Z\kern-.4emZ}}
\def\sZZ{\hbox{\zfonteight Z\kern-.4emZ}}
\def\RR{\hbox{\zfont  l\kern-.1emR}}

%
\def\boxit#1{\vbox{\hrule\hbox{\vrule\kern3pt
\vbox{\kern3pt#1\kern3pt}\kern3pt\vrule}\hrule}}
\newdimen\str

\def\fboxit#1#2{\vbox{\hrule height #1 \hbox{\vrule width #1
\kern3pt \vbox{\kern3pt#2\kern3pt}\kern3pt \vrule width #1 }
\hrule height #1 }}
\def\yboxit#1#2{\vbox{\hrule height #1 \hbox{\vrule width #1
\vbox{#2}\vrule width #1 }\hrule height #1 }}
\def\fillbox#1{\hbox to #1{\vbox to #1{\vfil}\hfil}}
\def\dotbox#1{\hbox to #1{\vbox to 8pt{\vfil}\hfil $\cdots$ \hfil}}

\def\yboxs{\yboxit{0.4pt}{\fillbox{5pt}}\hskip-0.4pt}

\def\tableaux#1{\vcenter{\offinterlineskip \halign{&\tabskip 0pt##\cr #1}}\ }
\def\cropen#1{\crcr\noalign{\vskip #1}}
\def\cry{\cropen{-0.4pt}}
\def\asym{\tableaux{\yboxs\cry\yboxs\cry}}
\def\fund{\kern+0.05em\tableaux{\yboxs\cry}\kern-.05em}
\def\afund{\kern+0.05em\overline{\tableaux{\yboxs\cry}\kern-.25em}\kern+.2em}

\def\Title#1#2{\nopagenumbers\abstractfont\hsize=\hstitle\rightline{#1}%
\vskip .45in\centerline{\titlefont #2}\abstractfont\vskip .5in\pageno=0}

\Title{
{
 \vbox{ \hbox{hep-th/9704246}
        \hbox{ILL--(TH)--97--1}
        \hbox{MIT--CTP--2613}
        \hbox{CERN--TH/97--07}
}}}
{\hsize =\hstitle \vbox{
        \centerline{Unity of Supersymmetry Breaking Models}
}}

\centerline{Robert G. Leigh\foot{e-mail: {\tt rgleigh@uiuc.edu}}}
{\it    \centerline{Department of Physics}
        \centerline{University of Illinois at Urbana-Champaign}
        \centerline{Urbana, IL 61801, USA}
}
\vglue .5cm
\centerline{Lisa Randall\foot{e-mail: {\tt lisa@ctptop.mit.edu}}}
{\it    \centerline{Center for Theoretical Physics}
                \centerline{Laboratory for Nuclear Science}
                \centerline{Massachusetts Institute of Technology}
                \centerline{Cambridge, MA 02139, USA}
}
\vglue .5cm
\centerline{Riccardo Rattazzi\foot{e-mail: {\tt rattazzi@mail.cern.ch}}}
{\it    \centerline{Theory Division, CERN}
                \centerline{CH-1211, Geneva 23, Switzerland}
}

\Date{}

\centerline{ABSTRACT}\bigskip
\baselineskip14pt\noindent
We examine the models with gauge group $U(1)^{k-1}\times\prod_{i=1}^k
SU(n_i)$, which are obtained from decomposing the supersymmetry breaking
model of Affleck, Dine and Seiberg containing an antisymmetric tensor
field. We note that all of these models are distinct vacua of a single
$SU(N)$ gauge theory with an adjoint superfield. The dynamics of this
model may be analyzed using the duality of Kutasov and Schwimmer and the
deconfinement trick of Berkooz. This analysis leads to a simple picture
for supersymmetry breaking for $k=2$, complementing that of previous
work. We examine the flat directions of these models, and give
straightforward criteria for lifting them, explaining the requisite
peculiar form of the superpotential. For all cases with $k>2$, the
duality argument fails to give supersymmetry breaking dynamics, and we
identify a class of problematic flat directions, which we term
$2m$-baryons. We study in some detail the requirements for lifting these
directions, and uncover some surprising facts regarding the relationship
between $R$-symmetry and supersymmetry breaking in models with several
gauge groups.

\baselineskip16pt
\vfill\break

\newsec{Introduction}
\seclab{\intro}

In previous papers\mnmods\ it was shown that a large class of models
could break supersymmetry. These models could be obtained from the
Affleck--Dine--Seiberg model \refs{\adsI,\meve,\ads}  
with an antisymmetric tensor and
antifundamentals simply by removing group generators\dnns\ and
decomposing the representations under a subgroup of the initial gauge
symmetry. The models considered had gauge group $SU(n_1)\times SU(n_2)
\times U(1)$; in what follows, we refer to these as the $n_1$--$n_2$--1
models. An interesting aspect is that the gauge group factors of the
resulting models  are in different phases. Nonetheless, models with
dynamical superpotentials, quantum confinement, smooth confinement, and
a more weakly coupled dual phase could all be shown to exhibit
supersymmetry breaking. Related work appears in Refs.
\refs{\ahem,\itdual,\ilsmods,\pts} and references therein.

  Although the gauge dynamics
of  the specific examples was different, there was a unifying
picture for the mechanism of dynamical supersymmetry breaking which
applied to all models. In all cases in which there was not initially a
dynamical superpotential for at least one of the gauge groups (that is a
superpotential generated by instantons or gaugino condensation which
would drive fields to large values), the dynamics of one of the gauge
groups was such that most flavors of the other gauge group were given
mass. In the low energy theory there was always a dynamical
superpotential for this second gauge group with the  few 
remaining light flavors.  In all cases, the low energy theory
contains a field  $\phi$ which is set to zero by the equation of motion of a
 composite field. This is inconsistent with another equation of motion, as
the dynamical superpotential removes the origin $\phi=0$.
This alone would
not suffice to break supersymmetry  if there
were to exist runaway directions.
Another common feature of all models considered in Refs. \mnmods\ was
that all flat directions could be lifted by a renormalizable
superpotential (though sometimes with a rather mysterious flavor
structure for the couplings, to be explained later). The common features of the supersymmetry breaking mechanism suggests the
existence of a more unified picture of all models. In this paper we show
that these models can be analyzed simultaneously by the introduction of
a massive adjoint field. The individual models can be understood as
particular vacua of the theory with the adjoint.

It might seem surprising that much can be learned from the theory with
an adjoint, which is notoriously difficult to understand. At present,
models with an adjoint are understood only in the presence of other
fields in the fundamental representation and in the presence of a
superpotential.\kutschw\ Our theories on the other hand contain an
antisymmetric tensor of each gauge group present. However, we can use
the deconfinement trick of Ref. \berkooz\ to interpret the theory as the
low energy limit of a confined $Sp(m)$ theory. The other remarkable
aspect is that  it is precisely the theory with a
superpotential considered by Kutasov that
is required for our analysis. In fact, the superpotential
$\Sigma^{k+1}$ is necessary in order to generate the vacua of the form
$U(1)^{k-1}\times\prod_{j=1}^{k} SU(n_j)$. Note that the models
considered previously\mnmods\ all correspond to $k=2$.

The very interesting result is that for the $k=2$ theories,  there exists a general and
very similar mechanism to that of Refs. \mnmods through
which supersymmetry breaking can be understood
simultaneously for all the examples. In particular, there is
a field whose equation of motion from the superpotential is set to zero
which is inconsistent with low energy dynamics of a dynamical
superpotential. However, here the dynamical superpotential is generated
by the $Sp$ dynamics, which is the only possible way all theories could
have been understood simultaneously! This is after many flat directions
are lifted through the combination of the superpotential and $SU$
duality.

This remarkably compact picture of all models suggests probing further
to the applicability of this picture to other models, the most obvious
generalization being the $k>2$ models. We find the dual picture is quite
different. Unlike the $k=2$ model, the dual theory is actually more
strongly coupled. Furthermore there are many more massless fields
charged under $Sp$ which remain in the low energy theory, so that the
analysis based on the 
$Sp$ dynamics is  inconclusive.

This leads to the  suspicion that the theories based on
$U(1)^{k-1}\times \prod_{j=1}^k SU(n_j),k>2,$ do not in fact break
supersymmetry. We argue based on explicit examples and a general
analysis of $R$-symmetry that these theories are unlikely to break
supersymmetry.  One key element is the emergence of new types of flat
direction, which we term $2m$-baryons (dibaryons in the $m=1$ case),
which cannot be lifted by the renormalizable superpotential, or indeed,
any superpotential which preserves an $R$-symmetry.

In this analysis, we learn some new facts about the role of
$R$-symmetry. In particular, many examples can be shown to break
supersymmetry with only an effective $R$-symmetry applying to a
low-energy version of the theory. Furthermore it is not always necessary
that the $R$-symmetry be anomaly-free under all gauge group factors.

In the next section we define our model and analyze the nonperturbative
dynamics
of the $k=2$ theory.  In section 3, we examine what can be learned from
the strongly coupled $k=3$ theory.  In section 4, we show that all flat
directions are lifted for the $k=2$ theories. Particular attention is
paid to the dibaryon and $2m$-baryonic flat directions. In section 5, we
consider in more detail some examples.  We argue based on these models
that we do not expect the $k>2$ vacua to give a supersymmetry breaking
vacuum.  We also clarify the role of $R$-symmetry in these models. We
conclude in the final section. An appendix completes the proof of the
lifting of flat directions in the $n_1$--$n_2$--$1$ models. A second appendix
shows that there are no dangerous flat directions in the deconfined
theory which would restore supersymmetry.

\newsec{Supersymmetry Breaking and Duality for $k=2$}
\seclab{\keqtwo}
\def\sc{{s_{k+1-j}\over j}}

We begin by considering an $SU(N)$ gauge theory, for odd $N$.
We take matter in the following representations of the gauge group:
\eqn\matter{
        \eqalign{
                \adj=& {\bf Adj}\cr
                A   =&\; \asym\cr
                \bF_I =&\; \afund,\;\;\; I=1,\ldots,(N-4).\cr
        }
}
This is just the matter content of Ref. \ads, with the addition of an
adjoint superfield. In general, we will have a superpotential for $\adj$
of the form
\eqn\sigsup{
W_\adj = \sum_{j=2}^{k+1} \sc \tr \adj^j .
}
For generic $s_j$, there are several discrete classical vacua, in which
$\adj$ is massive. First, there is a vacuum at $\vev\adj=0$, which has
the spectrum of the ADS model.\ads\ In addition, there are vacua with
gauge group $U(1)^{k-1}\times\prod_{j=1}^{k} SU(n_j)$ where $\sum
n_j=N$. For $k=2$, these models have been studied in Refs.
\refs{\mnmods}.

If  supersymmetry is to be broken, it is clear that couplings must be
added to \sigsup. For $k=2$, we consider a superpotential of the form
\eqn\super{
        W={1\over 2} m\tr \adj^2+{1\over3} s_0 \tr\adj^3
        +\l^{IJ} \bF_I A\bF_J +\tl^{IJ} \bF_I A\adj\bF_J
        +\ttl^{IJ}\bF_I\adj A\adj\bF_J
}
For each of the vacua mentioned above, this superpotential reduces, for
arbitrary $\lambda_i$, to the form necessary for supersymmetry breaking
\refs{\ads,\mnmods}.

In this paper, we will show that supersymmetry is broken dynamically in
the full theory \super, for all $m$. This may at first sight seem rather
strange; at $m=0$ it would appear that there is a
supersymmetry-preserving ground state at the origin, and there are
effectively too many flavors to lead to any non-perturbative effects. At
least in the absence of the superpotential, there are
effectively\foot{This is the quantity that appears in the
$\beta$-function, $N_{F,eff}= \sum_i T(R_i)$.} $N_{F,eff}=2N-3$; one
might expect then that the theory is in a non-Abelian Coulomb phase.
However, as we will show, and is clear from the supersymmetry breaking
analyses, the superpotential is quite relevant\foot{to our discussion,
as well as in the RG sense.}. We find that there is a useful dual
description in which the Yukawa terms are mass terms; the physics of
these is then clear. A full analysis of the dual theory leads to
supersymmetry breaking (in a way apparently independent of the existence
of the adjoint field) given certain minimum requirements on the Yukawa
couplings.
It is interesting that there is a link between the classical lifting
of flat directions and a simple dual non-perturbative phenomena (gaugino
condensation).
As mentioned in the introduction, the phenomena found
here is very similar to that of Refs. \refs{\mnmods}. In the present
case however, it is the duality of Refs. \refs{\kutschw} that is
used for the analysis.

\subsec{Deconfinement and Duality}

As we have indicated above, there is some indication that the idea that
the model \super\ is in the non-Abelian Coulomb phase may be fallacious.
The picture that we would like to advocate is that the Yukawa couplings
are, at the would-be infrared fixed point, relevant operators which
cause flow away to some new point (perhaps strong coupling, or another
fixed point). However, the description at hand is unable to make clear
the appropriate physics, as the Yukawa couplings are cubic, quartic
and quintic terms. This may not be the case in a dual description however.

\def\mup{\mu'}

A dual description of the theory \matter,\super\ is not available.
However, it is easy to derive one by combining the results of Refs.
\refs{\berkooz,\kutschw}. The technique is to `deconfine' the
antisymmetric tensor by introducing a new gauge group $Sp(m)$, for
$2m=N-3$, and matter (classified under $SU(N)\times Sp(m)$):
\eqn\mattersp{\eqalign{
        Y=\left( \fund,\fund\right)\cr
        Z=\left( {\bf 1},\fund\right)\cr
        \bP= \left( \afund, {\bf 1}\right)
        }}
and a superpotential $W=c (YZ)\bP$. The $Sp(m)$ group is confining;
the gauge invariant fields are $A=(YY)/\mup$ and $P=(YZ)/\mup$.
A superpotential
$Y^NZ\sim A^{m+1}P$ is generated and the tree-level superpotential
leads to a mass term for $P$ and $\bP$. Thus the low energy
theory is equivalent to the theory of an antisymmetric tensor without a
superpotential.

In the case at hand, we construct an $SU(N)\times Sp(m)$ theory
with matter
\eqn\matterdec{\eqalign{
                \adj=& \left( {\bf Adj}, {\bf 1}\right)\cr
                Y   =&\; \left( \fund , \fund \right)\cr
                Z  =&\; \left( {\bf 1}, \fund \right)\cr
                \bF_I =&\; \left( \afund , {\bf 1} \right)\;\;\;
                I=1,\ldots,(N-4)\cr
                \bP =&\; \left( \afund, {\bf 1} \right)
        }}
and the superpotential
\def\L{\Lambda}
\eqn\superdec{\eqalign{
        W=&{1\over2} m\;\tr\adj^2+{1\over3} s_0\;\tr\adj^3+ c\; Y\bP Z\cr
        &+{1\over\mup}\left\{ \l^{IJ} Y\bF_I\; Y\bF_J
        +\tl^{IJ} Y\bF_I\; Y\adj\bF_J
        +\ttl^{IJ} Y\adj\bF_I\; Y\adj\bF_J\right\}
        }
}
We are now in possession of a theory that we know how to dualize:
$SU(N)$ with an adjoint and $N_F=N-3$ flavors, and  $W \supset \tr\adj^3$.
  Treating $Sp(m)$ as a spectator, the dual theory
\kutschw\ will have gauge group $SU(\tilde N=N-6)\times Sp(m)$. The
matter includes $SU(\tilde N)$-singlet fields $M^a_I\sim Y^a\bF_I$,
$M^a_0\sim Y^a\bP$, $N^a_I\sim Y^a\adj\bF_I$ and $N^a_0\sim Y^a\adj\bP$,
as well as dual quarks $y^a$, $\brf^I$, $\bp$. There is also the adjoint
$\tadj$ and the spectator $Z$. The superpotential is\dkasns\
\eqn\superdual{\eqalign{
\tilde W=& -{1\over3}s_0\tr\tadj^3 -{1\over2}\tilde m\tr\tadj^2
+{1\over\mup}\tilde\l^{IJ} (M_IM_J)
+ {1\over\mup}\tilde\tl^{IJ} (M_I N_J)
+ {1\over\mup}\ttl^{IJ} (N_I N_J)
\cr
&+c(M_0 Z) +{s_0\over\mu^2}\left( M_I\brf^I\tadj y+M_0\bp\tadj y
+(\tilde b M_I+N_I)\brf^I y
+(\tilde b M_0+N_0)\bp y\right)
}}
where $SU(\tilde N)$ contractions are understood, and $Sp(m)$ contraactions
appear in
parentheses. Also, we have $\tilde\l=\l-b\tl+b^2\ttl$,
$\tilde\tl=\tl-2b\ttl$ and $b=m/2s_0$, $\tilde b=-bN_c/\tilde N_c$.
There is a subtlety here, as the field $M$ has dimension less than one
at the fixed point\kutschw, and thus should decouple in the infrared.
However, we also have a mass term for these fields, and it is
appropriate to integrate them out of the theory. The number of massive
modes clearly is dictated by the rank of the Yukawa coupling matrices.

To study this, collect the mesons together as $\Phi_I=(M_I,N_I)$, and
then the mass term can be written as $\Phi\cdot{\cal M}\cdot \Phi$, where
\eqn\bigyuk{
{\cal M}\equiv\pmatrix{\tilde\l & \tilde\tl \cr -\tilde\tl^T & \ttl}
}
The mesons also couple linearly to dual mesons $\Phi\cdot R$, where
$R=(\tilde b \brf y+\brf\tadj y , \brf y)$. If ${\cal M}$ has maximal
rank, we can integrate out all of the components of $\Phi$ and then only
$m-1$ flavors of $Sp(m)$ are left; this theory will generate a dynamical
superpotential. We find
\eqn\newsuper{
        \tilde W = \tilde W_{\tadj}(\tadj) + W_{irr}(y,\brf,\tadj)
        + {s_0\over\mu^2} \bp (N_0 y)
         +{\L_m^{m+2}\over \left( y^{\tilde N} N_0\right)^{1/2}}
}
where $\tilde W_{\tadj}$ is as in \superdual, and $W_{irr}$ is a
low-energy tree-level superpotential of the form
$R\cdot {\cal M}^{-1}\cdot R$.

It is convenient to define $\tilde A=yy/\tilde\mu$, $p=yN_0/\tilde\mu$,
 to rewrite the superpotential as
\eqn\superlow{
        \tilde W= \tilde W_{\tadj}(\tadj)
        +W_{irr}(A,\brf,\tadj)
        +{s_0\tilde\mu\over\mu^2}\ p\cdot\bp
        +{\L_m^{m+2}\over \tilde\mu^{(m-1)/2}
        \left(\tilde A^{m-2}p\right)^{1/2}}
}
The superpotential is thus very similar to that found in Refs.
\refs{\mnmods} for $SU(n_1)\times SU(n_2)\times U(1)$.  Therefore,
the F-terms for $p$ and $\bp$ are inconsistent.

We note however that we have not yet shown that flat directions are
absent. It is important to verify that such  flat directions do not
exist so that there is no runaway direction along which the potential
slopes to zero. To show that this does not occur, in a later section, we will
analyze the moduli space of the  individual vacua and argue that  they are 
compact. 
In all cases, the low energy theory
contains a field  which is set to zero by the equation of motion of a
 composite field. This is inconsistent with another equation of motion, as
the dynamical superpotential removes the origin.
 It is
encouraging that duality has given us the type of superpotential that we
would expect, similar to those of Refs. \refs{\mnmods}; apparently the
supersymmetry breaking mechanisms found in those vacua extend to the
full theory.

We have also obtained some insight into the necessary form of the Yukawa
couplings: they must have sufficient rank as to induce
non-perturbative dynamics for the $Sp(m)$ group, and thus induce
supersymmetry breaking. This is a necessary condition, but it is not
sufficient. There are further restrictions, which are related to the
comments above. To see this, let us explicitly invert the Yukawa matrix
\def\MiX{X}
\eqn\Minve{
{\cal M}^{-1}=\pmatrix{ \tilde\tl^{-T}\ttl \MiX^{-1} & -\MiX^{-T}\cr
\MiX^{-1} & \tilde\tl^{-1}\tilde\l\MiX^{-T}}
}
where $\MiX=\tilde\tl+\tilde\l\tilde\tl^{-T}\ttl$. An illuminating
example is the case $\ttl=\tilde\l=0$; then, for non-singular $\tl$,
${\cal M}$ is invertible. However, in this case, many of the dual meson
couplings are missing, and so we will expect flat directions to exist.
Thus in such a case, the $Sp$ group will indeed induce a non-zero
energy state, but there will presumably be runaways along
these classically flat directions. Thus, the full condition on the
$\lambda_i$-matrices involves a study of flat directions. In the
following sections, we will study this problem in more detail.

The form of the dual superpotential found here is somewhat intriguing.
Note that $W_{\tadj}$ and $W_{irr}$ apparently play no direct role in
the supersymmetry breaking (although they are the offspring of terms
necessary to lift flat directions). Both of these terms are of an
identical form to those of the electric theory. But also, the dual
Yukawa coupling scales, apart from shifts (due to a shift in the adjoint
field under duality) like the {\it inverse} of the electric Yukawa
coupling ${\cal M}$. This hints at some underlying duality in Yukawa
couplings.

\subsec{$N=7$}

Strictly speaking, the analysis above is valid for $N\geq 9$, as it is
only for these values for which there is a dual gauge group. For $N=7$,
the infrared physics is confining.\dkasns\ The physics of the confining
phase may be understood, as sketched in Ref. \dkasns, through duality.
In fact, the $N=7$ case (which would have a dual gauge group
``$SU(1)$'') may be read off from \superdual, with the replacement of
$\tadj=0$.\foot{In the following, we have fixed what we believe to be
minor typos in Ref. \dkasns (sec 6.2), related to setting $\tadj_s=0$,
instead of $\tadj=0$.} We thus obtain a confining superpotential
\eqn\confsuper{\eqalign{
\tilde W= {1\over\mup}\tilde\l^{IJ} (M_IM_J)
+ {1\over\mup}\tilde\tl^{IJ} (M_I N_J)
+ {1\over\mup}\ttl^{IJ} (N_I N_J)
+c(M_0 Z)\cr
+{s_0\over\mu^2}\left(
(\tilde b M_I+N_I)\brf^I y
+(\tilde b M_0+N_0)\bp y\right)
}}
In addition to this, there may be a non-perturbative contribution,
presumably of the form $\det\sum M$. The exact form of this
superpotential is not known; fortunately, we do not need to know its
detailed form for the present discussion. We need only that it does not
depend on the field $\bp$.

Again, we must integrate out massive meson singlet fields, and we then
obtain a result similar in form (up to the effect of the above-mentioned
non-perturbative contribution) to \newsuper. We see that the physics
relevant to supersymmetry breaking of the $SU(7)$ model is essentially
identical to that for $N\geq 9$.

\subsec{Duality  Without $\adj$}

The analysis of the flat directions which we will present applies to the
particular vacua of interest. We will not, apart from a few comments in
Section 4.4, analyze the flat directions including the $\Sigma$ field
which is integrated out of the theory. This is sufficient since it is
only the vacua which reproduce the supersymmetry breaking theories of
interest.  Nonetheless, it would be useful to verify that the $\Sigma$
field serves only as a device to generate the desired vacua, and does
not play an essential role in supersymmetry breaking dynamics. For this
reason, we show that an analysis very similar to that of the previous
section can be done once the vacuum is chosen. That is for a particular
\mnu\ model, one can deconfine the antisymmetric tensor via Sp
dynamics, eliminating the $\Sigma$ field altogether.

Consider the \mnu\ model which is parented by the $SU(N)$ ADS model with
$N=n_1+n_2$. The antisymmetric tensor $A$ decomposes into $A_1,\ A_2,\ T$.
Notice that the deconfinement trick we used before for $A$ can be used
to decompose its fragments as well. Doing so we have an $SU(n_1)\times
SU(n_2)\times Sp(m)$, with $2m= N-3$ and matter content
\eqn\noadmatter{\eqalign{
                Y_1   =&\; \left( \fund ,{\bf 1}, \fund \right)\cr
                Y_2   =&\; \left( {\bf 1},\fund , \fund \right)\cr
                Z  =&\; \left( {\bf 1}, \fund \right)\cr
             \bF_{ 1I} =&\; \left( \afund ,{\bf 1} , {\bf 1} \right)\;\;\;\cr
             \bF_{ 2I} =&\; \left({\bf 1}, \afund , {\bf 1} \right)\;\;\;
                I=1,\ldots,(N-4)\cr
                \bP_1 =&\; \left( \afund, {\bf 1} , {\bf 1} \right)\cr
                \bP_2 =&\; \left( {\bf 1}, \afund, {\bf 1} \right)\cr
        }}
The superpotential is
\eqn\supernoad{
        W=
        {1\over\mup}\gl^{IJ} Y_1\bF_{1I}\; Y_1\bF_{1J} +
        {1\over\mup}\gtl^{IJ} Y_2\bF_{2I}\; Y_2\bF_{2J} +
        {1\over\mup}\gttl^{IJ} Y_1\bF_{1I}\; Y_2\bF_{2J} +
        ( c_1\; Y_1\bP_1 + c_2\; Y_2\bP_2) Z
}
where the matrices $g_i$ can be written as linear combinations of the
$\lambda_i$ matrices of eq. \super. The discussion  parallels   the one
of the previous section.  The confined $Sp(m)$ theory is the \mnu\
model with the relevant low energy degrees of freedom   $A_1\sim Y_1^2$,
$A_2\sim Y_2^2$, $T\sim Y_1 Y_2$ and $P_{1,2}\sim Y_{1,2} Z$. A
superpotential $\sim A^{m+1} P$ (with obvious notation) is also
generated and the tree level superpotential leads to mass terms for
$P_{1,2}$ and $\bP_{1,2}$.  Let us choose $n_1>n_2$. Then for $5\leq
n_2\leq (N-1)/2$, both $SU(n_1)$ and $SU(n_2)$ admit an equivalent dual
description \NAD. The dual gauge group is $SU(\tnu)\times SU(\tnd)\times
U(1)$, with $\tilde n_{1,2}= N-3-n_{1,2}$. The matter includes
$SU(\tnu)\times SU(\tnd)\times U(1)$ singlet fields $M_{1I}^a \sim Y_1^a
\bF_{1I}$, $M_{2I}^a\sim Y_2^a\bF_{2I}$, $M_{10}^a\sim Y_1^a\bP_1$,
$M_{20}^a\sim Y_2^a\bP_2$ as well as dual quarks $y_{1,2}^a$, $\bar
f_{1,2}$, $\bar p_{1,2}$. The field $Z$ is a spectator. The
superpotential is now
\eqn\wnoadj{\eqalign{ \tilde W=&
 {1\over\mup}\gl^{IJ} M_{1I}\; M_{1J} +
        {1\over\mup}\gtl^{IJ} M_{2I}\; M_{2J} +
        {1\over\mup}\gttl^{IJ}M_{1I}\;M_{2J} +
        ( c_1\; M_{10} + c_2\; M_{20}) Z\cr
        &+ M_{1I}y_1\bar f_{1I}+M_{10}y_1\bar p_1+
        M_{2I}y_2\bar f_{2I}+M_{20}y_2\bar p_2\cr }
}

Now if both $\gttl$ and $\gttl+\gl^T\gttl^{-1}\gtl$ are
non-singular\foot{There is a clear analogy here with the matrix ${\cal
M}$ of Section 2.1.} we can integrate out all $M_{1I}$ and $M_{2J}$;
moreover also $Z$ and the combination $c_1M_{10}+c_2M_{20}$ pair up and
get a mass. Calling $N_0=c_2 M_{10} -c_1 M_{20}$, we are left with the
coupling $N_0(y_1\bar p_1+y_2\bar p_2)$. Then when $Sp(m)$ dynamics
generates a superpotential, we are lead to supersymmetry breaking, as
was the case in the theory with adjoint. In this case however we have a
proof that all flat directions are lifted and thus we can rigorously
conclude that there is a stable vacuum. A discussion of that appears in
Section 4 and in Appendix A. Notice that to reach our conclusion, we
have taken the limit where the dual gauge group $SU(\tnu)\times
SU(\tnd)\times U(1)$ gauge factor is weakly coupled and acts just as a
spectator: its only role is to lift flat directions at the classical
level.

\newsec{Supersymmetry Breaking and Duality for $k>2$}
\seclab{\kgttwo}

We now wish to attempt a construction similar to that of
previous sections for the cases $k> 2$.
We begin again with the matter \matter\ and a superpotential
of the general form
\eqn\superk{
        W=\sum_{j=2}^{k+1} \sc \tr \adj^j+
        \sum_{m,n=1}^{k}\lambda_{mn}^{IJ}\ \bF_I \adj^{m-1}A\adj^{n-1}\bF_J
 + \ldots}
Generically, the model has several classical vacua on the
Coulomb branch. To find these vacua, we consider solutions to the
$\Sigma$ equation of motion. This gives
\eqn\minimk{
\sum_{j=1}^{k} s_{k-j} x^j +\Delta =0
}
with the Lagrange multiplier $\Delta$ being determined by
the tracelessness of $\adj$,
$\Delta=-{1\over N_c}\sum_{j=1}^k s_{k-j}\tr\adj^j$.
A given vacuum will be determined by the number of each
solution to \minimk\ appearing in $\vev\adj$, that is, the number
of distinct eigenvalues.

We follow the construction of the $k=2$ case, deconfining
the antisymmetric tensor field in terms of an $Sp(m=N-3)$
gauge group. The resulting superpotential is
\eqn\superdeck{
        W=\sum_{j=2}^{k+1} \sc \tr \adj^j+\sum_{m,n=1}^{k}
                \lambda_{mn}^{IJ} (Y\adj^{m-1}\bF_I)(Y\adj^{n-1}\bF_J )
        + c\; Y\bP Z + \ldots
}

The dual theory\kutschw\ will have gauge group $SU(\tilde
N=(k-1)N-3k)\times Sp(m)$. The matter includes $SU(\tilde N)$-singlet
fields $(M_{(j)})^a_I\sim Y^a\adj^{j-1}\bF_I$, $(M_{(j)})^a_0\sim
Y^a\adj^{j-1}\bP$, as well as dual quarks $y^a$, $\brf^I$, $\bp$. There
is also the adjoint $\tadj$ and the spectator $Z$. The superpotential is
\eqn\superdual{\eqalign{
\tilde W= \sum_{j=2}^{k+1} {\tilde s_{k+1-j}\over j}\ \tr\tadj^{j}
+\sum_{j=1}^k \sum_{\ell} c_{\ell,j}
(M_{(j)})^a_0 \bp\tadj^{\ell}y^b {\cal J}_{ab}
+\sum_{j=1}^k \sum_{\ell} c_{\ell,j}
(M_{(j)})^a_I \brf^I\tadj^{\ell}y^b {\cal J}_{ab}\cr
+\sum_{m,n=1}^{k}\tilde{\lambda}_{mn}^{IJ} (M_{(m)})^a_I
(M_{(n)})^b_J {\cal J}_{ab}
+ c Z^a (M_{(1)})^b_0 {\cal J}_{ab}
}}
where $SU(\tilde N)$ contractions are understood, and ${\cal J}_{ab}$ is
the invariant tensor of $Sp(m)$. The $\tilde{\lambda}$ matrices are
linear combinations of the original Yukawa matrices, while $c_{\ell,j}$
are functions of the couplings appearing in the $\adj$-dependent part of
the superpotential.

The number of massive modes is again dictated by the rank of the Yukawa
couplings $\lambda_{mn}$. However, even if the number of massive modes
is maximal, there are still ${\tilde N +k-1\over2}=(2m+1)k/2-m-2$
flavors of $Sp(m)$ left massless (if $k$ is even).\foot{For odd $k$,
there is one massless mode in the $M$ mass matrix, and we get at least
$(2m+1)(k+1)/2-m-3$ flavors instead.} This is due essentially to the
presence of $y$: since $\tilde N$ grows with $k$, there are always many
flavors left; only in the case $k=2$ are there just $m-1$ flavors. The
$Sp(m)$ theory will not then generate a superpotential, and it is not
clear that supersymmetry is (or is not) broken. The most conservative
view is that it is not.

The analysis is clearly inconclusive. It is clear that for the $k>2$
models, $Sp$ dynamics in the dual theory will not in and of itself
suffice to prove supersymmetry breaking. In principle, the $SU$ dynamics
can be relevant, and lead to a supersymmetry breaking vacuum. Indeed
this is precisely what happens in the vacua that are dual to those
special electric vacua where $SU(N)$ is only broken to $SU(n_1)\times
SU(n_2)\times U(1)$. In other words, the $k>2$ case contains $k=2$
particular vacua. Consider for instance $k=3$ and focus on the vacua of
the magnetic theory where $SU(\tilde N=2N-6)\to SU(N-3)\times SU(\tilde
n_1)\times SU(\tilde n_2) \times U(1)^2$. The first $SU$ factor is
confining with a  quantum modified moduli space. When the confining
dynamics is accounted for, one reproduces the results of the $k=2$ case
considered previously. This analysis is analagous to that of Ref.
\kutschw.

In general however, it is not clear what role the $SU$ dynamics will play.
 These models therefore merit further investigation, which we do in a
later section. We will find that an essential distinction of the $k>2$
models is that there exist flat directions involving only the fragments
from the decomposition of the ADS adjoint field which are not lifted by
the cubic superpotential. It appears to be impossible to lift these
directions without introducing a supersymmetric minimum.

\newsec{$k=2$ Flat Directions }
\seclab{\flatdir}

Let us now return to the case $k=2$, and show that there are no flat
directions. We begin with a discussion of the individual vacua with
$SU(n_1)\times SU(n_2)\times U(1)$ gauge group. Following this, we
return to the full theory with $\adj$ in Section 4.4.

\subsec{Cubic  Invariants}

We study the \mnu\ model with a generic cubic superpotential. We will
derive a set of necessary requirements as well as a set of sufficient
ones that the Yukawa matrices must satisfy in order to classically lift
all flat directions. Rank maximality will turn out to be a necessary
requirement, without which there are unlifted flat directions. Other
requirements on the orientation in flavour space and on the eigenvalues
of Yukawa matrices will turn out to be sufficient to give a simple proof
that all flat directions are lifted.
The interesting result is that the Yukawa
matrices which satisfy all our requirements clearly represent
a set of ``non-zero measure'' in the space of couplings, {\it i.e.},
they represent a generic choice of couplings. In other words,
flat directions are not lifted only at special points in the space of
Yukawa couplings. However specifying what all these points are, {\it i.e.},
giving a set of necessary {\it and} sufficient requirements for lifting
flat directions, seems to require considerably more effort, for which there
is no apparent motivation. On the contrary, our analysis makes clear that
the somewhat mysterious choice of Yukawa matrices which seemed to be needed
in Refs. \mnmods\ corresponds to just one particular point in the vast
set of matrices which satisfy our sufficient requirements.
In the following discussion, we will make clear
which points in parameter space that we avoid. In this section, and in
Appendix A, we will prove that the cubic flat directions are all lifted.
In the following subsection, we will do the same for those corresponding
to the higher order invariants.

We denote by $A_{1,2}$ and $T$ the antisymmetric and mixed tensors
respectively, while $\fo= (\overline{n}_1,1)$ and $\ft=
(1,\overline{n}_2)$. As a convention, we will take $n_1 > n_2$.
The cubic invariants are
\eqn\inva{
\xo{ij}=\fo^i A_1 \fo^j,\quad
\xt{IJ}=\ft^I A_2 \ft^J,\quad
M^{Ij}=\ft^I T \fo^j.
}
(where $I,i=1,\ldots, n$, for $n=n_1+n_2-4$.)
The cubic superpotential may be written\foot{We have redefined the
couplings $g_i$; compare to the confined form of eq. \supernoad.}
\eqn\cubic{
W=g_{ij}\xo{ij}+f_{IJ}\xt{IJ}+\d{Ij}M^{Ij}
}
where $g_{ij}$, $f_{IJ}$ are antisymmetric matrices. We will show below
that a necessary condition for the classical lifting of all flat
directions is that $g,\,f,\,\delta$ be of maximal rank. In particular
$\delta$ must be invertible, so that by rotations and rescaling it
can be taken equal to the identity matrix $\delta ={\bf 1}$.

Let us study the F-term constraints given by the above superpotential.
By contracting the equations of motion to form invariants, we get two
classes of constraints, respectively linear and quadratic. The linear
ones are the following.
\eqn\lone{\eqalign {
\fo^k\partial_{\fo^i} W &= 2 g_{ij} \xo{kj}+\d{Ii} M^{Ik}\cr
&\ \to\ -2g X_1 + M=0\cr}
}
and
\eqn\ltwo{\eqalign {
\ft^K\partial_{\ft^I} W &= 2 f_{IJ} \xt{KJ}+\d{Ii} M^{Ki}\cr
&\ \to\ -2f X_2 + M^T=0\cr}
}
while from $A_1\partial_{A_1}$, $A_2\partial_{A_2}$ and $T\partial_T$
we get
\eqn\trace{
\tr (gX_1)=\tr (f X_2)= \tr (M)=0.
}
In the second lines of eqs. \lone\ and \ltwo, we have written the
expressions in matrix form.

There are many quadratic constraints; examples are
\eqn\mxone{
(T\fo^k)(A_1\fo^i)\partial_T W=\d{Ij}\xo{ij}M^{Ik}\ \to\ \xo{}M=0
}
\eqn\mxtwo{
(T\ft^K)(A_2\ft^L)\partial_T W=\d{Ij}\xt{LI}M^{Kj}\ \to\ -M\xt{}=0
}
One can easily show that,  when $\delta$ is nonsingular,
given eqs. \lone--\mxtwo,  the remaining quadratic constraints
are redundant.

It is easy to show that if the rank maximality is not satisfied the full
set of
equations admits non zero solutions. For instance if $\delta _{11}=0$ in
the diagonal basis, we  have that $M_{11}$ is totally unconstrained. And
similarly for $X_{1,2}$ entries when $g,f$ are not of maximal rank. So
we will from now on assume maximal rank for these matrices.

Since $n$ is odd, the generic antisymmetric matrices $g$ and $f$ will
have rank $n-1$. By a change of basis which leaves $\d{}$ invariant we
can always put one of them, say $g$, in the form
\eqn\gblockt{
g=\pmatrix{g'&0\cr 0&0\cr}
}
where $g'$ is an invertible $(n-1)\times (n-1)$ matrix. The matrix
$f$ will be of the form
\eqn\fblockt{
f=\pmatrix{f'&\rho\cr -\rho^T&0\cr}
}
where $\rho$ is an $n-1$ vector, and $f'$ is $(n-1)\times (n-1)$.
Now, $n-1$ is even so that it is clear that $f'$ will be singular only
at special points in parameter space. Then since we do not lose
much of parameter space and since it makes the discussion simpler
we will assume that $f'$ is invertible. In Appendix A we show that
it is sufficient to impose some additional mild requirements on $f'$ and $g'$
in order to easily conclude that all cubic flat directions are
lifted. We stress once more that the resulting space of matrices
is still of non-zero measure, consistent with what one may
call a principle of ``genericity''.
 Among the special points which
are removed from this space  will be the point $f'\propto (g')^{-1}$. The
removal of this point is indeed necessary, as it was shown in \refs{\mnmods}
that there are unlifted flat directions.
 Moreover the superpotentials of Refs. \refs{\mnmods}, in
particular, are simple examples which satisfy these requirements.

The genericity requirements derived in Appendix A place no restriction
on the vector $\rho$ in eq. \fblockt. Notice that for a  generic
$\rho\neq 0$, the superpotential preserves no non-anomalous R-symmetry.
An anomaly-free $R$ symmetry for these models is flavor dependent; a
generic superpotential breaks the $R$ symmetries.

\subsec{Higher order invariants}

We now consider invariants with more than three fields. These involve
$\epsilon$ and $\bar \epsilon$ tensors for each group. By $U(1)$
invariance and use of Fierz identities there are two classes of such
invariants. There are antibaryons, involving just $\bF_{1,2}$ contracted
to $\bar \epsilon_1\bar \epsilon_2$. These objects only exist for
$n_2\geq 4$, so that the smallest model which has these D-flat
directions is 5--4--1. The other class of invariants are baryonic, which
involve fields contracted to $(\epsilon _1\epsilon_2)^p$. This class may
be further divided into two subclasses, one involving matter fields
$\bar F_{1,2}$ and one whose elements are made purely of $A_{1,2}$ and
$T$. It turns out that elements in this second subclass, which we will
refer to as $2m$-baryons (or in the simplest case, dibaryons) vanish
identically. Since their vanishing is a peculiarity of $k=2$ models, we
will discuss them in more detail in the next section. In this section,
we study the antibaryons and the baryons which involve $\bF_{1,2}$.

The lifting of antibaryons is a consequence of $\partial _{A_{1,2}}W=0$,
together with $f$ and $g$ being of maximal rank. Indeed let us assume
that there remains a flat direction which has non-zero overlap with an
antibaryon. By combined flavor and gauge rotations we can go to a basis
where the vacuum expectation values are
\eqn\antivev{
\eqalign{ (\bF_1)^\alpha_i &= v_{1}^\alpha \delta^\alpha_i
\; 1\leq i\leq n_1\quad \quad (\bF_1)_i^\alpha=0 \; i> n_1 \cr
(\bF_2)^A_I &= v_{2}^A \delta^A_I
\; 1\leq I\leq n_2\quad \quad (\bF_2)_I^A=0 \; I> n_2 \cr}
}
where $\alpha$ and $A$ are gauge indices. The e.o.m. of $A_{1,2}$ are then
\eqn\aflat{
\eqalign {\partial _{A_{1\alpha\beta}}W &=(\bF_1)^\alpha_ig^{ij}
(\bF_1)^\beta_j=v_1^\alpha v_1^\beta g^{\alpha\beta}=0 \quad 1\leq
\alpha,\beta\leq n_1\cr
\partial _{A_{2AB}}W &=(\bF_2)^A_If^{IJ}
(\bF_2)^B_J=v_2^A v_2^B f^{AB}=0 \quad 1\leq
A,B\leq n_2\cr}
}
with no sum over $\alpha,\beta$ and $A,B$. The necessary condition for
the F-flatness of the antibaryon is the existence of a null $n_1\times
n_1$ submatrix in $g$ and of a null $n_2\times n_2$ submatrix in $f$.
This however conflicts with $g$ being of maximal rank (recall $n_1>n_2$
and $g$ is $(n_2+n_1-4)\times (n_1+n_2-4)$). Indeed it can be easily shown
that the existence of a $n_1\times n_1$ null submatrix bounds the rank
of $g$ to be $\leq (n_1+n_2-5)-(n_1-n_2+3)$, which is strictly smaller
than the maximal rank $n_1+n_2-5$. We conclude that all antibaryons have
to vanish on the equations of motion.

Now consider the baryons which involve at least one power of $\bar
F_{1,2}$. We will show that they may be reduced, by the F-term
constraints, to products of operators which either vanish trivially, or
are proportional to $2m$-baryon operators. The crucial remark here comes
by contemplating the F-term constraints of matter fields suitably
contracted to give such invariants. They are
\eqn\rone{ (P_1)\partial_{\bar F_1} W= (P_1)(T\ft)+(P_1)(A_1\fo)=0}
\eqn\rtwo{(P_2)\partial_{\ft} W= (P_2)(T\fo)+(P_2)(A_2 \ft)=0}
where $P_{1}$  and $P_2$ are combinations of fields and $\epsilon$'s
with, respectively, the same quantum numbers as $\fo$ and $\ft$. Now,
since the Yukawa matrix for $\fo T\ft$ is invertible, the above
equations tell us that each operator $O_1=P_1T\ft$, $O_2=P_2T\fo$ is
equal by the equations of motion to a combination of objects with one
less power of $T$. Heuristically, we can write this result as
\eqn\descend{
O_1=O_1\times ({A_1\fo\over T\ft})=O_1 \times r_1
\quad \quad
O_2=O_2\times ({A_2\ft\over T\fo})=O_2 \times r_2
}
where, for example, $\times r_1$ refers to the substitution in an
operator of $T \ft$ with $A_1\fo$. This result may now be
iterated, until one gets on the right hand side an operator which does
not have {\it enough} powers of $T$, and thus vanishes trivially. It
then follows that the original $O_{1,2}$, as well as all the
intermediate operators in the chain, must vanish by the equations of
motion.

Thus all operators vanish by equations of motion, provided they can be
written in the form $(\dots)T\ft$ (or $(\dots) T\fo$). It can be shown
by fierzing that any operator can be brought to  a form where only $T$'s
interpolate between $\epsilon$'s, while each $A$ has either both indices
contracted to the same $\epsilon$ or one contracted to $\bar F$.
Consider then a situation in which one $\epsilon$ is contracted to
$T$'s, $A_1$'s and to $(A_1\fo)$'s \eqn\fierz{(T^p)_K (A_1^q)_L
[(A_1\fo)^r]_M\epsilon^{KLM}} where $K, L, M$ are appropriate groups of
indices. By fierzing the gauge index of one $\fo$ into the $\epsilon$,
we  get a combination of the same operator plus one in which an $\fo$ is
contracted to a $T$. The original object is then of the form $(P) (T
\fo)$. The same conclusion is obtained by considering $A_2$'s and
$A_2\ft$'s instead. The only other possibility is that no $T$ is
contracted to an $\epsilon$ together with $(A_1\fo)$ or $(A_2\ft)$. 
Then either the operator  has a factor which is proportional to a
$2m$-baryon, or there is an $\epsilon$ for $SU(n_1)$ which is only
contracted to $A_1$'s, where we assume odd $n_1$ for the sake of the argument.
 The latter case vanishes trivially, while the former will
be  shown to vanish in the next section. We have thus succeeded in eliminating
flat directions involving $\bar F$'s.

Before moving on to $2m$-baryons, let us see how the above discussion
works for the simplest chain of invariants, that with only one power of
$\bar F$. These are given by (odd $n_1$)
\eqn\invariants{\eqalign {I_k&=A_2^{{n_2\over 2}-k}A_1^{{n_1+1\over 2}-k}
T^{2k}\fo=
I_0\times ({T^2\over{A_1A_2}})^k=I_0(r_1r_2)^{-k}\quad \quad k=0,\dots,
k_{\rm max}\cr
J_m&=A_2^{{n_2\over 2}-m}T^{2m +1}A_1^{{n_1-1\over 2}-m}\ft=
J_0\times({T^2\over A_1A_2})^m=J_0\times (r_1r_2)^{-m}
\quad\quad m=0,\dots,m_{\rm max}\cr}}
where for $n_1>n_2$ we have $(k_{\rm max},m_{\rm max})=(n_2/2,n_2/2-1)$,
while for $n_1<n_2$ we have $(k_{\rm max},m_{\rm
max})=(n_1/2-1/2,n_1/2+1/2)$.

Notice that $I_0\equiv 0$, and the above operators form the chain
\eqn\chain{
I_0\buildrel r_1 \over \longleftarrow J_0
\buildrel r_2 \over \longleftarrow I_1\, \cdots\, J_{\rm max}
\buildrel r_2 \over \longleftarrow I_{\rm max}
}
where by multiplying repeatedly by $r_{1,2}$ we move on the chain from
$I_{\rm max}$ down to $I_0$ (we consider here $n_1>n_2$). Now the
equations of motion \descend\  simply become \eqn\zerochain{0\equiv
I_0=J_0=I_1=J_1=\,\cdots\,=J_{\rm max}=I_{\rm max}} so that all the
operators are zero and the corresponding flat directions are lifted.

\subsec{2m-baryons}

In this section we discuss the baryonic invariants that are made only
of pieces of the antisymmetric tensor of the original ADS model and
not the antifundamentals.
The main result is that these objects  vanish for the
\mnu\ models ($k=2$) while they exist, and are important, for the
models with $k>2$. For this reason, this discussion
applies to general $k$. The  gauge group is
$$U(1)^{k-1}\times\prod_{i=1}^k SU(n_i)$$
where $\sum_i n_i=N$ and the antisymmetric tensor breaks down into:
\eqn\matter{
        \eqalign{
        A_i   =&\; ({\bf 1,\ldots,1},\asym_i,{\bf 1,\ldots,1})\;\;\;i=1,
        \ldots,k\cr
        T_{ij} =&\; ({\bf 1,\ldots,1},\fund_i,{\bf 1,\ldots,1},
        \fund_j,{\bf 1,\ldots,1})\;\;\;i,j=1,\ldots,k\;\;i\not = j\cr  }
}

It is simple to see algebraically, since all gaugeable $U(1)$'s are gauged,
that the most general baryonic operator can be written in the form
\eqn\barygen{
B_{\{m_i,p_{ij}\}}= \prod_i A_i^{m_i} T_{ij}^{p_{ij}}
}
with
\eqn\integers{2m_i+\sum_j p_{ij}=nn_i}
for each $i$, where $n$ is the number of $\epsilon$'s (for each group).
There is the possibility that there is more than one contraction for
each case, so there should be an additional label. Moreover, by summing
eq. \integers\ over $i$, since $N=\sum_i n_i$ is odd, we find that $n$
is even. We then conclude that any baryon has an even number $n=2 m$ of
$\epsilon$ tensors for each gauge group factor, so it is indeed a
$2m$-baryon (i.e., the $U(1)$ quantum numbers of the operator are all
$2m$.).

Let us study the case $k=2$ first. To see that no $2m$-baryonic
invariants exist, it is simplest to explore directly the required
$D$-term constraints. We require
\eqn\fle{2A_1^\dagger A_1 +{}^TT^\dagger {}^TT= c_1 1.}
\eqn\flo{2A_2^\dagger A_2 +T^\dagger T= c_2 1.}
\eqn\flu{2n_2 \tr(A_1^\dagger A_1)-2n_1\tr(A_2^\dagger A_2) +(n_2-n_1)
\tr(T^\dagger T)=0}
Now, by diagonalizing $A_1$ and $A_2$, eq. \fle\ tells us that
$\rk(T)=\rk({}^TT^\dagger {}^TT)$ is either odd or zero, while eq. \flo\
tells us that $\rk(T)=\rk(T^\dagger T)$ is even (zero included). This,
for consistency, requires $T$ to be of zero rank, {\it i.e.} $T=0$. Now
eq. \fle\ can only be solved for $A_1=0$, so that by eq. \flu, $A_2$ has
to also be zero.

However, once we consider $k>2$, one can show that $2m$-baryonic flat
directions exist. Indeed consider the $k=3$ case with gauge group
$SU(n_1)\times SU(n_2)\times SU(n_3)\times U(1)^2$. The antisymmetric
tensor of $SU(N)$ decomposes as
\eqn\kappa{A=\pmatrix{A_1&T_{12}&T_{13}\cr *&A_2&T_{23}\cr *&*&A_3}}
with notation as in eqs. \matter. Let us assume $n_1$ and $n_2$ are
even. Then for example, the following direction is flat:
\eqn\tiss{
T_{12}=\pmatrix{
                0&\cdots&0\cr
        .&\cdots&.\cr
        0&\cdots&x/{\sqrt {2}}\cr
        0&\cdots&0}\quad
T_{23}=\pmatrix{
                0&\cdots&0\cr
                .&\cdots& .\cr
                0&\cdots&x/\sqrt{2}\cr
                0&\cdots & 0 }
}
\eqn\tis{
T_{13}=\pmatrix{
                0&\cdots&0\cr
                .&\cdots&.\cr
                0&\cdots&x/\sqrt {2}} \quad \quad
}
with
\eqn\antis{\eqalign{
A_1&=x\ {\rm diag}(\sigma,\dots,\sigma,\sigma/\sqrt {2})\cr
A_2&=x\ {\rm diag}(\sigma,\dots,\sigma,\sigma/\sqrt {2})\cr
A_3&=x\ {\rm diag}(\sigma,\dots,\sigma,0)\cr
}}
The $U(1)$ D-terms are important here. Indeed there exist $SU(n_1)\times
SU(n_2)$ flat directions involving only $A_1$, $A_2$ and $T_{12}$ which
are lifted by the $U(1)$'s. Similar flat directions can also be found in
the case in which all three $n_i$ are odd. Obviously the above is enough
to assess the existence of these objects for the case $k\geq 3$, since
it  can be obtained for higher $k$ by group reduction. We will consider
these flat directions more fully in Section 5.

\subsec{Flat directions in the full model with $\adj$}

Finally we would like to make a comment on the associated flat
directions in the original theory with the adjoint $\ad$, {\it i.e.}
those involving just $A$ and $\ad$. Since the $\bar F$ fields are not
excited, $F$-flatness is just given by
\eqn\fsigma{\eqalign{
{\partial W\over \partial \ad}&=c_k\ad^k+\cdots+c_1 \ad+b 1=0\cr
b&=-{1\over N}\sum_j c_j \tr(\ad^j)
}}
On the other hand, D-flatness corresponds to
\eqn\dasy{
A^\dagger A +\ad^\dagger \ad -\ad\ad^\dagger=c 1.
}
Now it can be easily shown that for $k=2$ eq. \fsigma\ constrains $\ad$
to be such that $D_{\ad}=\ad^\dagger \ad -\ad\ad^\dagger$ has even rank,
and thus eq. \dasy\ has only the trivial solution $A=0$, $D_\ad=0$. This
does not happen at $k>2$. The peculiarity of $k=2$ is more easily seen
by considering the simple potential $W=\tr \ad^{k+1}$. Now at $k=2$ the
equation of motion is just $\ad^2=0$, which implies that $\ad^\dagger
\ad$ and $\ad \ad^\dagger$ are orthogonal to each other. Thus we have
$\rk(D_\ad)=2 \rk(\ad \ad^\dagger)$. Now already at $k=3$ we have
$\ad^3=0$ and we cannot conclude much on the rank of $D_\ad$, which can
now in fact be odd. Take for instance $SU(3)$ and
\eqn\highrank{
\ad= u\pmatrix{0&1&0\cr0&0&{\sqrt 2}\cr0&0&0}
}
Now we have
\eqn\hyper{
[\ad^\dagger,\ad]=|u|^2\pmatrix{1&0&0\cr0&1&0\cr0&0&-2.}
}
Indeed the similar ways in which $k=2$ stands out as a special case in the
above discussion and in duality may be worth further investigation.

\newsec{ Examples and the Role of $R$-Symmetry}

In the previous sections, we have found various pieces of evidence
consistent with supersymmetry breaking for the $k=2$ models. On the
other hand, the $k>2$ models behave quite differently; the duality
construction failed to show supersymmetry breaking, and furthermore, we have
been able to identify unlifted flat directions. In this section, we
consider these models further, paying particularly close attention to
the flat directions and $R$-symmetry, both in the general case and in
specific examples.  We will consider adding additional operators to the
superpotential, in an attempt to classically lift the $2m$-baryons. In
the examples, it is not possible to lift all $2m$-baryonic flat
directions (when they exist) and preserve an $R$-symmetry. When they are
not lifted, there is a gauge symmetry breaking minimum (a runaway
direction) in which the dynamics responsible for breaking supersymmetry
does not occur.  On the other hand, when the $2m$-baryons are lifted
and no $R$-symmetry is preserved,
there exists a supersymmetric solution to the equations of motion, which
would be expected on the basis of the argument of Nelson and
Seiberg.\aenns

However, the role of $R$-symmetry is not always clear. In examples where
the superpotential is nongeneric, an $R$-symmetry is not
essential\aenns. A trivial example of this is given by the superpotential
$W=\phi_1+ \phi_2^2 +\phi_2^3$ (or, more generally, by a
superpotential
which decomposes as the sum of two terms $W_{1,2}$ which depend on
different fields and such that $W_1$ is $R$-symmetric while $W_2$ is not).
In some cases, indeed, the lack of genericity of the microscopic
superpotential may lead to an {\it effective} $R$-symmetry
of some low-energy version of the theory.  In the above mentioned case
this is what happens after $\phi_2$ is integrated out.
If on the other hand, the
superpotential is generic to a sufficient degree, {\it i.e.} it involves
enough independent operators,
one would not expect the symmetries of the
low-energy theory to include an $R$-symmetry not present in the original
theory. In a subsection below, we reanalyze the $4-3-1$ model with a
general cubic superpotential in a particular limit in which there exists a
hierarchy of strong mass scales. With such a general dimension three
superpotential there is no quantum $R$-symmetry. Nonetheless
the superpotential of the low energy theory has an $R$-symmetry
and supersymmetry is broken, consistent with the analysis of Ref. \aenns.
It seems that this low energy symmetry originates because of
the simple, non generic, {\it just} cubic form of the microscopic $W$.
 The remaining examples we
consider, corresponding to $k=3$ models, will not have this luxury;
higher dimensional operators are required to lift dangerous flat
directions. No $R$-symmetry of these theories can be identified when the
flat directions are lifted and the models will not break supersymmetry.

Before proceeding to specific examples, we give a general discussion. It
is useful for us to first focus on  a  particular subset of $D$-flat
directions. First, there are the dimension-three flat directions which,
as in the analogue eq. \inva, we call $X$ and $M$. The second set of
flat directions we distinguish are the $2m-$baryons, $Y$, discussed in
detail above. In particular we focus on the dibaryons. These have
dimension $N$, the size of the original ADS $SU(N)$ gauge group from
which the model was derived. It is important to notice that the dibaryon
operators, unlike the other higher dimension operators, cannot be lifted
by the renormalizable superpotential terms, since each term involves at
least two $\bar{F}$-type fields. Therefore, additional operators must be
present in the superpotential in order to lift these flat directions
classically.  The dibaryons can be lifted either by including the
dibaryon in the superpotential, or by including a higher order invariant
with a single $\bar{F}$ factor.


We now give an argument that it is impossible to put in all $X$, $M$,
and $Y$ operators while maintaining an anomaly-free $R$-symmetry. It is
readily checked that if $M$ and $X$ operators are present in the
superpotential, the dibaryon operator carries $R$-charge
\eqn\ryprime{R_{Y}=2N-2\sum_i n_i \bar{F_i}} where we have used the
field name to represent its $R$-charge. On the other hand, the anomaly
cancellation condition for $SU(n_i)$ requires that
\eqn\ranomi{2(N-n_i-2)=\sum_j n_j \bar{F_j}.} Clearly one cannot satisfy
this constraint for all $n_i$. Furthermore, the $R$-charge of $Y$ will
not be two, and so the dibaryon will break this $R$-symmetry. If one
looks for flavor-dependent R-symmetries, one finds that the only flavor
symmetries allowed by the maximal rank condition on the Yukawas that
could mix with $R$ are non-anomalous. The same conclusions are reached
if we try to lift the dibaryon by including a higher order invariant
with only one $\bar F$.

The above argument shows that one cannot consistently include all
dimension-three gauge invariant operators in the superpotential,
maintain an anomaly-free $R$ symmetry, and lift the dibaryon
classically. This might suggest that one cannot construct a
supersymmetry breaking model based on this gauge and field content. This
argument is suggestive but inconclusive. In fact it is unclear that any
of the $R$ symmetry requirements above are essential for each gauge
group. First of all, it is not clear that all $M$ operators need be
included. Omitting these operators introduces new flat directions which
break one of the $SU(n_i)$ symmetries. However, if it is assumed that
another $SU(n_j)$ gauge group is strong first, the resulting flat
direction can be lifted quantum mechanically. So although it is clear
that at least one type of $M$ operator must be included (where type
refers to which type of $A$ and $\bar{F}$ operator it involves), it is
not clear that {\it all} are necessary.

Second, it is not really clear that the dibaryonic operators need to be
included. Although it is obvious that the dibaryonic operators are not
lifted by the dimension three superpotential operators, they can
conceivably be lifted quantum mechanically.\refs{\ahem,\shirman}

The third point is perhaps the most subtle. It is not necessary to
incorporate all anomaly constraints, even for non-Abelian groups. This
goes against conventional wisdom, but we give two arguments why this can
be the case. There are two arguments in the literature concerning the
special role of $R$-symmetry in supersymmetry breaking. The first
argument\ads\ is that if there is a spontaneously broken global symmetry
and no flat directions the theory is likely to break supersymmetry since
the massless pseudoscalar has no massless scalar partner. The second
argument\aenns\ shows that if there is a generic superpotential and an
$R$-symmetry, there are more equations which must be satisfied for a
supersymmetric minimum than unknowns, and therefore the theory will not
have a supersymmetric minimum.

Now let us consider both of these arguments in turn. Suppose we have a
classical symmetry which is anomalous with respect to a particular gauge
group factor, but that factor does not contribute a superpotential in
the electric phase. A U(1) gauge symmetry is an example of such a
factor, but a gauge group in the non-Abelian Coulomb phase or free
magnetic phase might also have this property, depending on how it is
perturbed. In this case, the K\"{a}hler potential (and higher
derivatives terms) alone would violate
the global symmetry, typically
through higher dimensional operators which depend on
the dynamical scale $\Lambda$ of the theory. The axion would only
get a mass from the K\"{a}hler potential {\it after} supersymmetry is
broken. So if one is asking whether there can be a supersymmetric
minimum, the fact that the global symmetry is anomalous is
irrelevant.

Now let us consider the Nelson-Seiberg argument. Since this argument
only depends on the equations of motion, one can look directly at
the superpotential to see how the anomaly constraint enters. It enters
precisely as we have considered above; that is, if there exists an
operator generated by strong dynamics present in the superpotential, it
is one of the terms considered when analyzing the equations of motion.
Clearly if there is an instanton-generated term, for example, it should
be consistent with the $R$-symmetry. This is guaranteed if the
$R$-symmetry is not anomalous with respect to the gauge group being
considered. However, if the $R$-symmetry is anomalous with respect to a
factor which does not generate an operator in the superpotential, it is
clearly irrelevant to the Nelson-Seiberg argument. The conclusion is
that even for non-Abelian gauge groups, one does not necessarily need to
require an anomaly free $R$-symmetry.

For the above three reasons, it is very difficult to make completely
generic statements about all models, since it might be that there exists
a particularly clever choice of operators such that some flat directions
remain or the $R$-symmetry is anomalous, but nonetheless there exists a
supersymmetry breaking  minimum. For example,  in the \mnu\ models, it
is not essential to preserve an exact $R$-symmetry.  The  analysis of
the $4-3-1$ model in the next section demonstrates that there is an
effective $R$-symmetry in some low-energy version of the theory which is
sufficient to guarantee supersymmetry breaking.  
The $4-3-1$ model
represents an example where the $SU$ dynamics cannot be
neglected, {\it i.e.} the scales of both gauge groups
appear in the effective superpotential.
 On the other hand, the proof of dynamical 
supersymmetry
breaking in the deconfined version of the \mnu\  models without an
adjoint relied solely on the $Sp$ dynamics of the dual phase. Therefore,
the $SU(\tilde n_1)\times SU(\tilde n_2)$ can be taken very weak without
spoiling the susy breaking dynamics, which is all determined by $Sp(m)$.
It is easy to verify in this case that there is an $R$-symmetry which is
anomaly-free with respect to the $Sp$ gauge group (but anomalous with
respect to $SU$).

The $n_1-n_2-n_3$ models on the other hand will probably not break
supersymmetry. In these models, there are generally two possible
formulations of the superpotential. In one, the dibaryon direction is
not lifted. An $R$-symmetry is preserved, but the flat direction leads
to a supersymmetric minimum at infinity. In the second formulation, the
dibaryon is included, but there is no remaining $R$-symmetry and there
is a supersymmetric minimum at finite field value. This will be
demonstrated in the  examples which follow. As for the general
$n_1-n_2-n_3$ models in the dual phase we may argue in the following
way. The $Sp$ dynamics alone in the dual theory is not enough to
generate a superpotential. Therefore, the dynamics of  at least one of
the $SU$ factors  must be relevant if supersymmetry is to be broken.
This suggests that if supersymmetry is to be broken, for the purpose of
classifying $R$-symmetries one must also impose some $SU$ anomaly
constraint, and not only the $Sp$ anomaly constraint. By imposing the
constraint for both $Sp$ and $SU(\tilde n_1)$ we find, again, that in
the presence of the most general cubic superpotential all dibaryons have
charge $R_Y=2(k-1)N-6k-4 -4\tilde n_1$. This expression can equal 2 only
for special groups.

\subsec{Example1: The $4-3-1$ Model}

In this section we  reanalyze this model in order to illustrate the role
of an R-symmetry, whether exact or accidental. It is known that this
model does break supersymmetry, unlike the models considered later in
this section. However, unlike Ref. \mnmods, we will consider the model
with a generic cubic superpotential, which will not in general preserve
an $R$-symmetry. Nonetheless the model can be shown to break
supersymmetry with an $R$ symmetry  of a low-energy effective
superpotential as we now discuss.

Since this model is completely confining, its behavior is distinctive
when compared to the larger \mnu\ cases where at least one of the group
factors is, at least naively, in a non-Abelian Coulomb phase. In this
case the confining superpotential, derived in Ref. \mnmods, involves
both strong scales. Thus we expect that the anomaly constraints of both
groups are relevant to R-symmetry considerations. As we will describe
below the situation is however more subtle. Let us analyze  it in
detail. The field content under $SU(4)\times SU(3)\times U(1)$ is just
given by
\eqn\content{
A(6,1)_6,\,\ \
\bar Q(1,\bar 3)_{-8},\,\ \
T(4,3)_{-1},\,\ \
\bF_I(\bar 4,1)_{-3},\ \
\bar Q_i(1,\bar 3)_4
}
where $i,I=1,2,3$ are the flavor indices. The most general cubic tree
level superpotential of the model can be written as
\eqn\notr{
W=g^{12}\bar Q \bar Q_1\bar Q_2 +f^{12} A \bF_1\bF_2 +
\lambda^{iJ} T\bar Q_i\bF_J}
where $\lambda$ is a rank 3 matrix, which is in general non-diagonal. We
have already proven that the above $W$ lifts all flat directions. For
diagonal $\lambda$ there is  an anomaly-free R-symmetry \mnmods\ under
which the fields have charges $A(0)$, $\bF_3(0)$, $\bF_{1,2}(1)$, $\bar
Q_{1,2}(5/3)$, $\bar Q_3(8/3)$, $\bar Q(-4/3)$, $T(-2/3)$. This symmetry
is indeed anomalous under the $U(1)$ gauge group, but this anomaly will
not play a role as there is no strong dynamics associated with the
$U(1)$. However for general $\lambda$ there is no quantum $R$-symmetry.
More generally we notice that whenever one of $g^{12}$, $f^{12}$,
$\Lambda_4$, $\Lambda_3$ vanishes or when $\lambda$ is diagonal there
exists a non-anomalous $R$-symmetry. Incidentally we notice also that in
the case of generic Yukawa matrices the only non-anomalous $U(1)$ is
indeed the gauged one. Nonetheless the model with general couplings, and
no $R$-symmetry, does break supersymmetry. For this model the full confining
superpotential is known.\mnmods\ Nonetheless we find it instructive to
consider the limit $\Lambda_3\gg \Lambda_4$ and study the effective
theory below the scale of $SU(3)$ confinement. We want to show that,
while the original microscopic theory does not preserve an $R$-symmetry,
the effective low-energy one does indeed have an accidental $R$ in the
superpotential. The original breaking appears only in the K\"{a}hler
potential and in higher derivatives terms. This effective $R$-symmetry
plays the usual role\aenns\ in supersymmetry breaking.

The $SU(4)$ gauge theory below the scale of $SU(3)$ confinement contains
4 flavors ($F_i$, $\bar F_{I}$), one antisymmetric tensor $A$, and 4
singlets $\bar b^i$.  In terms of the original fields we have $F_i\sim
T\bar Q_i$, $\bar b^i=\epsilon^{ijk}\bar Q\bar Q_j\bar Q_k$ for
$i=1,2,3$ and $\bar F_4 \sim T^3$, $F_4\sim T\bar Q$, $\bar b^4\sim \bar
Q_1\bar Q_2\bar Q_3$. The superpotential of the confined theory can be
shown to be just the tree level plus the confining terms
\eqn\wconfined{W= g^{12} \bar b_3 + f^{12} A\bar F_1 \bar F_2
+\lambda^{Ij}\bar F_I F_j+ {1\over \La{3}^5}(\bar F_4 F_i \bar b^i-{\rm
det} F_i)}
where  the indices of $\lambda$, the tree level Yukawa, run from 1 to 3.
The original Yukawa couplings $\la{}$ are now mass terms for three of
the four $SU(4)$ flavors, so that it is appropriate to integrate them
out. Indeed the remaining light fields are just spectators of this
decoupling and the low energy $W$ is just obtained from \wconfined\ by
setting the massive fields to zero
\eqn\weffective{W_{eff}= g^{12} \bar b_3+
\bar F_4 f_4\bar b^4}
This result is easily derived by redefining $\bar F_3$ in such a way
that $\la{}$ has the form
\eqn\massterm{\lambda^{Ij}=\pmatrix{\la{11}&\la{12}&\la{13}\cr
\la{21}&\la{22}&\la{23}\cr 0&0&\la{33}}.}
In this basis, the e.o.m. of $\bar F_3$ and $F_{1,2}$ imply $F_3=0$ and
$\bF_{1,2}\propto \bF_4$. Then both  $A \bar F_1\bar F_2$ and ${\rm det}
F$ vanish by the e.o.m., while the linearity of $W$ in $F_{1,2}$ leads
to the vanishing of the other terms involving the heavy fields and to
the simple result \weffective. Moreover the scale of the low energy
theory is just given by\foot{The dimensions do not match here since we
have not canonically normalized composites.} $\tilde \La{4}^{10}={\rm
det \lambda}\ \La{4}^8\La{3}^5$. The low energy $SU(4)$ dynamics
generates a superpotential from gaugino condensation and the full low
energy $W_{eff}$ will be \eqn\wfull{W_{eff}=g^{12}\bar b^3+{1\over
\La{3}^5}M_{44}\bar b^4+\left ({ \tilde \La{4}^{10}\over
PfA\,M_{44}}\right )^{1\over 2}} where $M_{44}=\bar F_4 F_4$. This
expression preserves an $R$-symmetry. The reason for this is that a
number of sources of explicit $R$-breaking have decoupled. In
particular, $W_{eff}$ does not depend at all on $f$ while the dependence
on $\La{4}$ and $\la{}$ is all coming via ${\rm det}
(\la{})\La{4}^8$. Notice that, for any $\la{}$, the latter expression is
neutral under the $R$ symmetry defined below eq. \notr. Eq. \wfull\ is
exact; it can be derived from the full effective superpotential of Ref.
\mnmods\ by integrating out heavy mesons along generic $Pf A\not = 0$.
Alternatively one could derive it by considering the most
general low-energy effective $W$ under the simple assumption that
$SU(3)$ confinement gives a mass to $\bar F_{1,2,3}$ and $F_{1,2,3}$.
One can then use the constraints from the $SU(3)\times SU(3)$ flavor
symmetry of the original theory, under which the fields transform as
$\bar Q_i (3,1)$,
$\bF_I(1,3)$, $g(3,1)$, $f(1,3)$ and $\la{}(\bar 3, \bar 3)$. Flavor
symmetry and holomorphy then constrain the Yukawa couplings to appear in
the low energy theory only via the three expressions $I_1=g^{ij}\bar
b^k\epsilon_{ijk}$, $I_2={\rm det}(\la{})$ and $I_3=g^{ij}f^{IJ}\la{kL}
\epsilon_{ijk}\epsilon_{IJL}$. Notice the field independent $I_{2,3}$
are indeed $R$ preserving: so holomorphy and flavor symmetry alone are
already enough to infer $R$ invariance of the low energy $W$ ! (The fact
that the last expression above does not appear in $W$ is not even
necessary for our purpose.)

The occurrence of such an accidental $R$-symmetry is indeed analogous to
what happens in the $SU(2)$ model of ref. \iss. We can trace the origin
of $R$ to the specific form of the microscopic superpotential. Had we
added additional quintic or higher order invariants to the original $W$,
the equations of motions would not have lead to the vanishing of all the
terms involving heavy fields, and we would not be left with an R
symmetry.

Notice that there should be no exactly massless $R$-axion associated to
the breaking of $R$, as the K\"{a}hler potential and the terms with
higher covariant derivatives do break $R$. It is also clear, that, since
an exact $R$ symmetry is recovered in the limit $\La{4}\to 0$, the mass
of a possible $R$-axion scales like $\La{3}(\La{4}/\La{3})^p$ with
$p>0$. For comparable scales $\La{3}\sim \La{4}$ there should be no
approximately massless Goldstone boson. This should be compared to other
models where supersymmetry is broken without an $R$-symmetry. In Ref.
\aenns, the 3-2-1 model with the addition of one flavor $s,\,\bar s$ of
$SU(3)$ is considered. This model, with the most general renormalizable
superpotential, supports no $R$-symmetry. Nonetheless the model breaks
supersymmetry for any finite value of the mass term $m \bar s s$. At
$m=0$ supersymmetry is restored. Refs. \iss\pts\ also give models which
do break supersymmetry without an $R$ symmetry. However in both these
cases, there are non-renormalizable operators which stabilize some flat
direction. The resulting axion mass, though suppressed by some power of
$1/\mpk$, may still be large enough to suppress axion production in
stars. Models with supersymmetry breaking induced by higher dimensional
operators need however to have a strong dynamics much above the
``minimal'' $10^{4-5}$ GeV.

The common feature of the previous supersymmetry breaking models without
an $R$ symmetry is that there is a dimensionful parameter in the tree
level Lagrangian, $m$ in Ref. \aenns\ and $1/\mpk$ in Ref. \iss\pts. The
4-3-1 model which we have just described, on the other hand, has all its
scales generated by dimensional transmutation, and still the $R$ axion
gets a mass. This seems to be the first example of this type. Indeed
this is somehow similar to $R$ breaking via the addition of an $R$-color
gauge group, which was discussed, though without  explicit examples, in
Ref. \aenns. R-color is assumed weaker than the dynamics responsible for
supersymmetry breaking, and its only purpose is to make $R$  anomalous.
In the limit in which $\La{3}\gg \La{4}$, the role of $SU(4)$ is similar
to that of an $R$-color factor.

To conclude we comment on this result. The main point is that an
effective $R$, limited to the low energy superpotential, can result from
a microscopic $W$ which is general enough to lift all flat directions
and break $R$, though not {\it completely} generic. A cubic $W$ at $k=2$
seems to have this remarkable property. Notice that had we studied 4-3-1
with comparable scales for the two groups there would not have been an
$R$ symmetric low energy theory. In that case we would have concluded
that supersymmetry is broken, even without $R$, due to the
non-genericity of the full superpotential. It is interesting that by
moving the scales we can go from a picture where $W$ is non-generic and
there is no $R$ to one in which a low energy $W$ is  generic but
also $R$-symmetric. We reiterate that the addition of enough higher
dimensional operators would of course eventually restore supersymmetry.
The nice thing about $k=2$ models, which can make them appealing in
applications, is that this ``non-genericity'' just results from
renormalizability. In this sense it is natural. Supersymmetry is broken
at $k=2$ just because we can lift all flat directions with a very
limited set of operators. The $k>2$ models do not have this critical behaviour,
since the lifting of flat directions requires too many operators. On the
other hand at $k=1$, {\it i.e.} the original ADS models, the cubic
superpotential does not break $R$ and there is an axion.

\subsec{Example 2: The $3-1-1$ Model}

The field content of this model is obtained by decomposing the ADS
model based on gauge group $SU(5)$ into
its components under an $SU(3)\times U(1)\times U(1)$ subgroup.
The antisymmetric tensor decomposes as
\eqn\qbarqqs{
A\to \bar{Q}(-4/3,0)\oplus Q_1(1/3,1)\oplus Q_2(1/3,-1) \oplus
S_3(2,0)}
where there is only one $A$ type field, $\bar{Q}$, and there are three
$T$ type fields. The antifundamental decomposes as
\eqn\qbarsost{
\bar{F}\to \bar{Q_1}(2/3,0)\oplus S_1(-1,-1)\oplus S_2(-1,1)}
We define the flat directions
\eqn\xeqn{
X_1=S_1 (\bar{Q}_1Q_2), \quad
X_2=S_2 (\bar{Q}_1 Q_1), \quad
X_3=S_1 S_2 S_3
}
\eqn\ypeqn{
Y=(\bar{Q} Q_1) (\bar{Q} Q_2) S_3
}
\eqn\zeqn{
Z_1=(\bar{Q} Q_1) (\bar{Q}_1 Q_2),  \quad
Z_2=(\bar{Q} Q_2) (\bar{Q}_1 Q_1)
}
And the superpotential is of the form
\eqn\supthroo{W=X_1+X_2+X_3+Y+{\Lambda_3^7\over Z_1-Z_2}}
It is readily checked that this superpotential does not support an
$R$-symmetry and that there are consistent supersymmetric solutions to the
equations of motion.

We now ask whether  it is possible to  preserve an $R$-symmetry by
omitting superpotential terms (in this example, the instanton generated
term requires that an $R$-symmetry be nonanomalous with respect to
$SU(3)$).   One can consider removing one or more of the $X$ operators
and/or the $Y$ operator from the superpotential.  First consider
removing $X_1$ or $X_2$. We see that in this case, we will not lift the
operators $Z_1$ or $Z_2$ which would  then be a runaway direction.  It
can readily be seen that without the  $Y$ operator, the equations of
motion require that $X_1$ and $X_2$  vanish, which would require $Z_1$
and $Z_2$ to diverge. So  we conclude we cannot omit any of the above
superpotential operators (without adding something else) if we are  to
get a minimum at finite expectation value. This is not surprising as it
is expected that the $X$ operators should have maximal rank in order to
avoid dangerous flat directions. This minimum preserves supersymmetry.

There exist other possible superpotentials to lift the flat directions
of the theory. It can be checked that the other possible models work
similarly. One might note the similarity of this model to models with an
antisymmetric tensor for even $N$.  These models break supersymmetry at
any finite field value if the operator  $A^{N/2}$ is omitted from the
superpotential. However, without the operator, there is a runaway
direction and a supersymmetric minimum exists at infinity. With the
inclusion of the operator, the $R$-symmetry is destroyed and there is a
supersymmetric minimum at finite field value. In the $3-1-1$ model, the
theory with the $Y$ operator has a supersymmetric minimum, while the
theory without it has a supersymmetric minimum at infinity. The presence
of the new flat direction $Y$, not present in the $n_1-n_2$ models, is
critical to the analysis of supersymmetry breaking.

\subsec{Example 3:  The $2-2-1$ Model}

We next consider an example with two non-Abelian gauge groups in the
decomposition of $SU(5)$, namely $SU(2)_L\times SU(2)_R\times U(1)$, where
we have labelled the two $SU(2)$'s for convenience of notation. The field
content for this decomposition is
\eqn\antidec{A \to Q_L(-3,1)\oplus Q_R(-3,-1)\oplus V(2,0)
\oplus S_L(2,2)\oplus S_R(2,-2)}
and
\eqn\antifdec{\bar{F}\to S(4,0)\oplus F_L(-1,-1) \oplus F_R(-1,1)}
There are many possible flat directions in this model. The
dimension-three flat directions are
$X=F_L V F_R$, $X_L=S F_L Q_L$, $X_R=S F_R Q_R$,
the dimension four flat directions are
$Z_L=\det V\cdot (F_L Q_L)$, $Z_R=\det V\cdot (F_R Q_R)$,
$T=S\cdot Q_L V Q_R$, $R_L=S_L S_R F_L Q_L$, $R_R=S_L\cdot S_R F_R Q_R$,
$W_L=S_L\cdot  F_L V Q_R$, $W_R=S_R F_R V Q_L$,
and the dimension-five invariants are $Y_1=\det V\cdot Q_L V Q_R$ and
$Y_2=S_L S_R Q_L V Q_R$. There are constraints among these directions
but they do not affect the following analysis. The flat directions can
be lifted by the superpotential
\eqn\suptwtwo{
W=X+X_L+X_R+Y_1+Y_2.
}
This superpotential does not preserve an anomaly-free $R$-symmetry
however. In fact this theory does not break supersymmetry. It is
interesting to see this explicitly. The strong dynamics associated with
the product of $SU(2)$ groups with this field content was worked out in Ref.
\pts. The superpotential which results is
\eqn\wstrong{\eqalign{
W_{eff}=A\left (B_L B_R u-B_L \Lambda_R^4-B_R
\Lambda_L^4-M_{11}M_{22}M_{12}M_{21}\right)\cr
+M_{11} +S B_L +S B_R +{u M_{22}}+S_L S_R M_{22}
}}
where the bound states of $SU(2)$ are $B_L=F_L Q_L$, $B_R=F_R Q_R$,
$u=det V$, $M_{11}=F_L V F_R(=X)$, $M_{12}=F_L V Q_R$, $M_{21}=Q_L V F_R$,
$M_{22}=Q_L V Q_R$. One can check explicitly that the equations of
motion can be solved. The situation is very similar to the previous example.
Without the $Y$ operators, there would have been
runaway directions at which supersymmetry is restored. With
the inclusion of the $Y$ operators, there is a supersymmetric
minimum at finite field value.

Again, we are left with the question of whether or not one can lift
dangerous flat directions while preserving an $R$-symmetry. Let us first
assume that we include the dimension-three superpotential as above. We
also impose the anomaly constraints since both SU(2)'s are confining and
their associated $\Lambda$'s appear explicitly in the superpotential.
One can then check that there is a two parameter family of
$R$-symmetries, under which the charges are $F_R=-2F_L+4+Q_R$,
$V=-2+F_L-Q_R$, $S=-2+2F_L-2Q_R$, $Q_L=-3F_L+4+2Q_R$ (the $R$-charges of
$S_L$ and $S_R$ are also free). One can then derive the charges of the
flat directions. Most flat directions are already lifted. The directions
$T$ and $Y_1$ are not. However they have $R$-charge $0$ and $-2$,
respectively. In order to preserve an $R$-symmetry in the
superpotential, $T$ and $Y_1$ can only appear multiplying a flat
direction which has been lifted. Therefore one cannot lift the $Y_1$
operator consistent with an $R$-symmetry and the presence of the
remaining terms in the superpotential. We conclude that there is no
superpotential which will preserve an $R$-symmetry and have a
supersymmetry  breaking minimum, at least with this choice of tree-level
superpotential.

There are however other combinations of tree level superpotential
which might be tried.  If either $X$, or both $X_L$ and $X_R$
are removed, there would be flat directions along which both $SU(2)$'s
are Higgsed, and there would be no dynamical supersymmetry
breaking. If only $X_L$ is removed, $Z_L$ must be lifted
or else once again both $SU(2)$'s would be Higgsed. So the operator
$Z_L$ should be included in the superpotential.  In this case, $X_L$
is still not lifted and can obtain a nonzero expectation value. If
this were the case, there is an $SU(2)_R$ theory with one flavor
due to the nonzero vev's of the fields in $X_L$.  One can then
check that without $R_L$ or $Y_L$ there would be a supersymmetry
breaking vacuum at finite field value, but of course there are runaway
directions. Once one of these operators is included, there is
a supersymmetric vacuum.

We conclude there is no model with this field content which
breaks supersymmetry.

\subsec{Example 4:  The $3-2-2-1-1$ Model}

The last model we consider explicitly is based on decomposing
$SU(7)$ to $SU(3)\times SU(2)_L \times SU(2)_R \times U(1)\times U(1)'$.
The fields decompose as
\eqn\thtwtw{\eqalign{
A \to \bar{Q}(\bar{3},1,1)_{(0,8/3)}
\oplus S_L(1,1,1)_{(2,-2)}\oplus S_R(1,1,1)_{(-2,-2)} \cr
\oplus T_L(3,2,1)_{(1,1/3)}\oplus T_R(3,1,2)_{(-1,1/3)}
\oplus V(1,2,2)_{(0,-2)}
}}
and
\eqn\barf{
\bar{F_7^I}\to \bar{F}^I(\bar{3},1,1)_{(0  ,-4/3)}
\oplus f_L^I(1,2,1)_{(-1,1)}\oplus f_R^I(1,1,2)_{(1,1)}
}
We can once again ask whether it is possible to lift the dibaryonic flat
directions as well as all other potentially dangerous flat directions.

Again there are numerous possibilities. For example, we can include all
dimension-three operators $T_L \bar{F} f_L$, $T_R \bar{F} f_R$, $f_L V
f_R$ since, as we have seen, in addition to lifting some of the
dimension-three flat directions, these operators permit most
higher-dimension operators to be lifted as well. Since the
$SU(3)$ group is confining,
and below this scale (after integrating out massive flavors),
the  $SU(2)$ groups are confining (and it
is their dynamics which is relevant), we need to impose the
$R$-anomaly constraints associated with these three gauge groups  in
order to consider   an $R$-symmetry which can serve as a useful guide to
supersymmetry breaking.

We now need to lift the dibaryon flat directions.  If the
$\bar{Q}\bar{F}\bar{F}$ operators are not present in the superpotential,
there are flat directions along which the $SU(3)$ symmetry is Higgsed.
The theory then reduces to the $2-2-1-1$ model (with some additional
singlets) and will not break supersymmetry. It is readily seen that one
cannot include the dibaryon $V^2 T_L^2 T_R^2 \bar{Q}$ in the
superpotential while preserving an $R$-symmetry. The only possibility is
to lift the dibaryon operators with lower dimension invariants. However,
to include such an operator, $\bar{F}T_R^2 T_L^2$ would prevent
including an operator $\bar{Q}\bar{F}\bar{F}$ in the superpotential, if
an $R$-symmetry is to be preserved.   It is interesting to note that if
we do not preserve an $R$-symmetry, and write a generic potential to
lift all flat directions, $SU(3)$ confines and again one is reduced to
the $2-2-1-1$ model so supersymmetry is not broken.

\subsec{$n_1$--$n_2$--$n_3$ Models in Dual Phase}

The models considered above all had a gauge group in the confining
phase. It is interesting to ask why the mechanism for breaking
supersymmetry which was common to the $m-n$ models no longer applies.
Although in models with confinement, there can be other sources of
supersymmetry breaking, all the $m-n$ models that we studied could be
considered in the limit where first one the the groups got strong. This
resulted in a theory in which Yukawa couplings became mass terms, so
that there were sufficiently few flavors for the other gauge group that
a dynamical term ensued. Furthermore, as a remnant of the dynamics of
the first group, there was a mass term between some field which only
coupled in this term and another field which occurred in the dynamical
superpotential. Because this field was set to zero by the equations of
motion, there was no consistent solution to the equations of motion in a
theory without flat directions.

In the $n_1-n_2-n_3$ theories, several of the pieces are missing. First
of all, the field whose equation of motion set a field to zero now
multiplies the sum of two fields, so neither are necessarily vanishing.
Second, even after integrating out massive flavors, there can still be
so many flavors that neither group generates a dynamical potential.
Although it could be that the groups are in the non-Abelian Coulomb phase
and no definite conclusion can be made, there is no evidence that such a
theory will break supersymmetry.  Finally, it might be that one or the
other group has sufficiently few flavors to generate dynamically a
superpotential. But as in the explicit examples we have studied, we
expect that there will be a supersymmetric solution to the equations of
motion in a theory in which the dibaryons are lifted.

\newsec{Conclusions}

To summarize, we have found that the addition of an adjoint superfield
gives a compact way to investigate a large class of product group models.
For the $k=2$ models, supersymmetry breaking was understood in
the dual picture through the (deconfining) $Sp$ dynamics. For the $k>2$
models, the dual description uncovered no dynamics which  would lead
to supersymmetry breaking.

An interesting aspect of the analysis of the dual phase was that maximal
rank Yukawa matrices were required in order to reduce the number of
flavors sufficiently (for $k=2$) that there was a gaugino condensation
contribution to the superpotential. An explicit
analysis of flat directions yielded further insight into this
requirement. We found that
most maximal rank superpotentials can lift all flat directions for
$k=2$.

We have also argued that $k>2$ models, that is models with more than
three non-Abelian factors and/or more than one Abelian factor, do not
break supersymmetry. This can be attributed to flat directions which
could not be lifted while maintaining a sufficiently nongeneric
superpotential, or while maintaining an $R$-symmetry.

Searching for an anomaly-free $R$-symmetry can be an unreliable guide to
determining whether supersymmetry is broken. One often requires only an
effective $R$ symmetry, and not even that when the superpotential is
non-generic (the latter in agreement with Ref. \aenns). We have found in
particular that similarly to examples presented there in which massive
fields can be integrated out to produce an effective $R$-symmetry,
dynamically massive fields can be integrated out to do the same.  This
behaviour is found in the $4-3-1$ model with generic superpotential for
example, where the anomaly-free $R$-symmetry of the superpotential of
Ref. \mnmods\ was not in fact necessary. Furthermore, we argued that
anomaly-free $R$-symmetries are required only when the dynamics of the
associated gauge group is somehow reflected in the superpotential.
Otherwise, an anomalous $R$-symmetry is permitted, as it is in the case
of Abelian gauge factors.

It is clear that the tools we have developed should be useful in
exploring other models with higher rank tensors. For example, a model
with symmetric tensors can be treated the same way.

However, it is also clear that there is a lot about dynamical
supersymmetry breaking that we have yet to understand. Analyzing flat
directions is almost always difficult and subtle. Furthermore it is not
clear when these flat directions are dangerous, as they might be lifted
by strong dynamics.  Finally we have found $R$-symmetries
useful, but often inconclusive. Essentially the same $4-3-1$ model with
and without an $R$-symmetry breaks supersymmetry. On the other hand,
clearly the fact that we cannot include dibaryonic operators is related
to $R$-symmetry breaking. It would be worthwhile to have even stronger
tools for analyzing potential dynamical supersymmetry breaking theories.

\bigskip
\noindent {\bf Acknowledgements:} We thank Csaba Cs\'aki and Witold
Skiba for useful conversations and comments on the manuscript.
R. L. and R. R. thank MIT for hospitality, and L. R. thanks the theory group of
Rutgers University for its hospitality when this work was initiated.

\appendix{A}{}

In this appendix, we complete the proof of Section 4.1 that all cubic
invariants in the $n_1-n_2-1$ models are lifted. Recall the form of the
F-term constraints
\eqn\minimala{\eqalign {
M&=2g\xo{}=2\xt{} f\cr
M\xt{}&=\xo{}M=0\cr
\tr (M)&=0\cr}
}

As noted in the text,
since $n$ is odd, the generic antisymmetric matrices $g$ and $f$ will
have rank $n-1$. By a change of basis which leaves $\d{}$ invariant
we can always put one of them, say $g$, in the form
\eqn\gblock{
g=\pmatrix{g'&0\cr 0&0\cr}
}
where $g'$ is an invertible $(n-1)\times (n-1)$ matrix. The matrix
$f$ will be of the form
\eqn\fblock{
f=\pmatrix{f'&\rho\cr -\rho^T&0\cr}
}
where $\rho$ is an $n-1$ vector, and $f'$ is $(n-1)\times (n-1)$. As explained
in section 4,
by genericity, we will assume $f'$ to be invertible as well. To study
eqs. \minimala\ in this basis it is useful to decompose also $X_{1,2}$ as
\eqn\xblock{
X_{1,2}=\pmatrix{X'_{1,2}&v_{1,2}\cr -v^T_{1,2}&0\cr}
}
The first of eqs. \minimala\ then gives
\eqn\oneblock{
gX_1=   \pmatrix{g'X_1'& g'v_1\cr 0& 0\cr}
        =       \pmatrix{X_2'f'-v_2\rho^T & X_2'\rho \cr -v_2^T f'& -v_2^T
\rho}
        = X_2 f
}
which implies
\eqn\twoblocka{v_2=0}
\eqn\twoblockb{v_1=(g')^{-1} X_2' \rho}
\eqn\twoblockc{X_1'=(g')^{-1} X_2' f'.}
Another useful set of constraints are
\eqn\xoneblocksq{
X_1 g X_1=\pmatrix {X_1'g'X_1'& X_1'g'v_1\cr
-v_1^Tg'X_1'&-v_1^Tg'v_1\cr}=0
}
\eqn\xtwoblocksq{
X_2 f X_2= \pmatrix {X_2'f'X_2' & 0\cr 0&0\cr}=0
}
which, together with eqs. \twoblocka--\twoblockc, give
\eqn\crucialone{X_1'g'X_1'=0}
\eqn\crucialtwo{X_2' f'X_2'=0}
In fact, eq. \crucialone\ is redundant: it may be deduced from eqs.
\twoblockc\ and \crucialtwo. It also important to keep in mind that
$X_1'$ is antisymmetric, so for example,
\eqn\antisym{(g')^{-1} X_2' f'= f' X_2' (g')^{-1}.}
The constraint given by this equation is crucial. Notice that for
the particular point $f'\propto (g')^{-1}$ it would be trivially
satisfied for any $X_2'$, leading to unlifted flat directions
\refs{\mnmods}. For
example, taking both matrices equal to the identity matrix,
which might have seemed the most obvious choice, will not work.
It is also useful to rearrange the above equations to deduce
\eqn\gminus{
X_2' (g')^{-1} X_2'=0.
}
We will now focus on the above two equations and show that there exists
a set of non zero measure of $f',\,g'$ for which they imply
 $X_1'=X_2'=0$. Then eqs. \twoblocka--\twoblockc\ and
\minimala\ allow us to conclude that the complete set of cubic invariants
$M$, $X_1$ and $X_2$ has to vanish by the equations of motion.

It is useful to notice that, while keeping the form
$M_{Ij}\sim \delta_{Ij}$, it is possible to make further rotations
and rescaling such that $g'$ and $f'$ in $[(n-1)/2] \times [(n-1)/2]$ blocks
have the form
\eqn\gfinal{ g'=\pmatrix { 0&1\cr -1& 0\cr}}
\eqn\ffinal{f' =\pmatrix {f_1& H\cr -H^T & f_2\cr}}
where $f_{1,2}= -f_{1,2}^T$.  A crucial assumption that we are going to
make here is that $H$ have $\ell=(n-1)/2$ distinct eigenvalues, so that it
can be diagonalized by a similarity transformation
\eqn\similarity{
H=U^{-1} d\ U \quad \quad d={\rm diag}(d_1,\dots,d_k)\quad
\quad d_i\not = d_j \quad i\not = j
}
  Finally, we make an additional
change of basis which simplifies the discussion; we define
\eqn\vbasis{
V=\pmatrix {U^{-1} & 0\cr 0& U^T\cr}
}
and make the redefinitions $X_2'\to (V^{-1})^T X_2' V^{-1}$, $f'\to
V f' V^T$, $(g')^{-1}\to V (g')^{-1} V^T$, which replace $H$ with $d$
while leaving $g'$ invariant.
By decomposing $X_2'$ in $\ell\times\ell$ blocks
\eqn\xtwofinal{
X_2'=\pmatrix{ b_1&\Delta\cr -\Delta^T &b_2\cr}
}
eq. \antisym\ reduces to
\eqn\commbone{ [d,b_2 ]=\Delta^T f_1-f_1 \Delta}
\eqn\commbtwo{[d,b_1]=\Delta f_2-f_2 \Delta^T}
\eqn\deltacomm{[\Delta^T,d]=b_2f_2-f_1b_1}
while eq. \gminus\ becomes
\eqn\bonetwo{b_1b_2-\Delta^2=0}
\eqn\bone{b_1\Delta^T+\Delta b_1=0}
\eqn\btwo{b_2 \Delta + \Delta ^Tb_2=0.}

The models studied in Refs. \refs{\mnmods} are a
particular example of the case $f_{1,2}=0$.
In this case, eqs. \commbone--\commbtwo\ imply
$b_{1,2}=0$, since the eigenvalues of $d$ are all non-degenerate.
Moreover, eq. \deltacomm\ constrains $\Delta$ to be diagonal,
so that by eq. \bonetwo\ we must have $\Delta=0$ and all cubic
invariants must then vanish.

In Ref. \mnmods, the superpotential that was chosen corresponds to
\eqn\hcyclic{H_{ij}= \delta_{{\rm mod}_k (i),{\rm mod}_k( j+1)}}
whose $\ell$ eigenvalues are all non degenerate and are
given by the $\ell$-th roots of the identity: $d_j=e^{i 2 \pi j/\ell}$.
We stress that the non-degeneracy of the $d$'s (eq. \similarity )
is the main reason why those particular
examples succeed in lifting all flat directions. For non-degenerate $d$'s,
eqs. \commbone--\deltacomm\ are maximally constraining. Indeed
in the limiting situation of $H=1$ and $f_{1,2}=0$ there would be flat
directions.

Notice that one easily gets the same result also in the case
in which just one of $f_{1,2}$, say $f_1$, vanishes while $f_2\not = 0$.
In order to make our case that flat directions are lifted at generic
points we should now consider the case where
 both $f$'s are non zero. This is however  more involved, and we have not
found a simple argument showing that $X_2=0$. It is however
possible to reach this conclusion for small enough,
but otherwise general, $f_1$ and $f_2$. For our purpose of showing
removal of flat directions over a set of non-zero measure of
the space of parameters this  suffices. On a case by case
basis it is  not  difficult to study the above equations for
any size of the $f$'s. It is straightforward to do that for the case
$\ell=2$, which corresponds to models generated by the $SU(9)$ ADS,
and the result is that $X_2=0$. The proof for the case $f_{1,2}\ll d_j$
is just done by ``perturbing'' the one we gave at $f_{1,2}=0$.
The off-diagonal entries in eqs. \commbone--\deltacomm\
allow one to solve for $b_{1,2}$ and for $\Delta_{ij}$, $i\not = j$,
in terms of $\{\Delta_{ll}\}$, which we will now simply
indicate with $\Delta_l$. In particular we have $b_{1,2}={\cal O}(f_{2,1})
\Delta$, while $\Delta_{ij}={\cal O}(f_1 f_2)\Delta$, for $i\not = j$.
It is then easy to see that the diagonal entries of eq. \bonetwo\
have the form
\eqn\deltasquare{\Delta_l^2 +B_{lmn}\Delta_m\Delta_n=0, \quad\quad
l=1,\dots, k}
 where $B_{lmn}$ is of order $f_1 f_2$.
For $B$ small enough, the above set of $\ell$ equations has only the solution
 $\Delta_l=0$. Indeed, if to the contrary, we assume
a non-zero solution exists, then consider
the equation for the largest $\Delta_{\rm max}$ in $\{\Delta_l\}$
\eqn\maximum{|\Delta_{\rm max}|^2=|B_{l_0mn}\Delta_m\Delta_n|<{\cal O}
(f_1f_2)|\Delta_{\rm max}|^2.}
This, for $f_1f_2$ small enough, has only the solution  $\Delta_{\rm max}=0$,
i.e. $\Delta_l=0$.
To conclude let us list the requirements on $f'$ for which we
have been able to prove complete lifting of flat directions:
\item {i)} $d_i\not =d_j$ for $i\not = j$ for the eigenvalues of $H$.
\item {ii)} $|f_1 f_2|\ll |d_i-d_j|$ for any $i\not = j$ and where on
the left-hand side we mean the product of any entry of $f_1$ with any entry of
 $f_2$.
While this is not the full space of matrices it is clearly a set of
non-zero measure, proving that a generic
cubic superpotential lifts all the cubic flat directions.
 Notice that
no $R$-symmetry is preserved with this superpotential when $\rho$ in eq.
\fblock\ is non zero.

\appendix{B}{}
\def\minimal{\lone--\mxtwo}

In this Appendix we briefly discuss the lifting of flat directions in
the deconfined theory. We will just consider the theory without adjoint
with tree level superpotential given by eq. \supernoad. By construction,
this theory reduces to the original \mnu\ model when $Sp(m)$ confines.
This means that at any finite point on the $Sp$ moduli space we can
integrate out the massive $P$, $\bar P$ and get a theory which has no
flat directions. The question remains whether there are flat directions
along which $Sp$ is higgsed on which  the potential slopes to infinity.
To answer this question we need to understand the space of classically
flat directions of the deconfined theory. The invariants are easily
obtained by forming $Sp$ mesons first. This procedure gives the
$A_{1,2}$ and $T$ of the \mnu\ theory, plus $P_{1,2}\sim Z Y_{1,2}$. The
latter are however set to be zero by the equations of motion of $\bar
P_{1,2}$. So the invariants that are not obviously vanishing are formed
from the fields of the \mnu\ theory plus $\bar P_{1,2}$. The analysis
goes through similar to those of Section 4 and Appendix A. The trilinear
and baryonic invariants are shown to vanish by the equations of motion
for maximal rank {\it generic} Yukawa matrices $g,f,\delta$. Indeed one
is reduced to eqs. \minimal\ for the cubic invariants, provided the
condition ${\rm Det}(\delta+4g{}^T \delta^{-1}f)\not = 0$ holds as well.
Notice that this is precisely the condition of maximal rank for the
meson mass matrix of the dual theory. The difference with respect to the
confined case is however that antibaryons are not lifted by the
classical equations of motion. This is because $Y_{1,2}$ couple
quadratically (rather than linearly) to matter and their equations of
motion certainly do not constrain the antibaryons at $Y_{1,2}=0$.
However, the only directions which can remain flat will not involve the
$Y_1$,$Y_2$, and $Z$ fields.  Therefore, any remaining  flat directions
do not Higgs the $Sp$ gauge group. In other words, the classical moduli
space of the deconfined theory consists just of all the antibaryons that
are formed from $\bF_1^i, \bar P_1$ and $\bF_2^I,\bar P_2$. Along any
such flat direction $SU(n_1)\times SU(n_2)\times U(1)$ is completely
higgsed, while $Sp(m)$ remains strongly coupled. Then we still expect
that the $Sp$ can be treated as confining and that the antibaryons are
indeed lifted by quantum effects. To see this explicitly we analyze the
theory far away along one such antibaryon $\bar B\sim \bar
\phi^{n_1+n_2}$, where $\bar\phi$ is the elementary field that makes up
$\bar B$. For the sake of the argument we may consider an antibaryon
which does not involve $\bar P_{1,2}$ so that by a flavor rotation the
field vev has the form
\eqn\anti{
\eqalign{ (\bF_1)^\alpha_i &= \bar \phi\delta^\alpha_i
\; 1\leq i\leq n_1\quad \quad (\bF_1)_i^\alpha=0 \; i> n_1 \cr
(\bF_2)^A_I &= \bar \phi \delta^A_I
\; 1\leq I\leq n_2\quad \quad (\bF_2)_I^A=0 \; I> n_2 \cr}
}
where $U(1)$ invariance fixes the vev to be the same for $\bF_1$ and
$\bF_2$. For large enough $\bar \phi$ the K\"{a}hler metric is flat in
$\bar \phi$. The original $Sp(m)$ theory has $m+2$ flavors. For maximal
rank Yukawa couplings, along an antibaryon, between $(n_1-n_2+3)/2$ and
$m+2$ flavors get massive. Remember that $n_1\geq n_2+1$, so that at
least two $Sp$ flavors get massive. Let us denote the massive $Sp$
fundamentals by $Q_A$ for $A=1,\dots,2p$ and the massles ones by $q_a$
for $a=1,\dots,2m+2-2p$. Moreover there are massless $Sp$ singlets $\bar
\psi_k$  coming from the antifundamentals of the broken $SU(n_1)\times
SU(n_2)$. The $\psi_k$ parameterize indeed the other unexcited
antibaryons, thus we take by definition $\psi_k=0$ on the antibaryon
$\bar B$. The classical superpotential in terms of these fields has the
form
\eqn\wclass{W=\bar \phi^2\Gamma_0^{AB}Q_AQ_B+\bar \phi\psi_k\Gamma_1^{kab}
q_aq_b+\psi_k\Gamma_2^{kab}q_aq_b+\dots}
where $\det \Gamma_0\not = 0$ gives mass to $p$ flavors of $Sp(m)$ and
the first two terms originate from the quartic terms in the original
$W$, while the third orginates from $\bar P_{1,2} ZY_{1,2}$. The dots
represent terms quadratic in $\psi$, terms of the type $\bar \phi \psi
Q_AQ_B$, and terms like $\bar \phi\psi Qq$, which are not relevant for
the following discussion where we focus on $\psi\sim 0$. Notice that the
property that $Y_{1,2}$ and $Z$ be zero by the classical equations of
motion have to be reproduced by the above superpotential at the point
$\psi=0$. This property is trivially satisfied for the massive flavors
$Q$ as $\Gamma_0$ is non singular. On the other hand the light mesons
$m_{ab}=q_aq_b$ have to be set to zero by the $\psi$ e.o.m.. In order
for this to hold true the fields
$\Psi^{ab}= \psi_k (\Gamma_1^{kab}+\Gamma_2^{kab}/\bar\phi)$,
for $a,b=1,\dots,2m+2-2p$ should be linearly independent so that
$\partial_{\Psi^{ab}}W=m_{ab}=0$. Now, by integrating the massive $Q$'s
out we get a low-energy $Sp$ dynamics which generates a superpotential,
described by the following $W_{eff}$
\eqn\quantumeff{W_{eff}\sim \left ({\bar \phi^{2p} \Lambda_{Sp}^{m+1}
\over Pf(m_{ab})}\right )^{1\over {p-1}} + \bar\phi \Psi^{ab}
m_{ab} }
Notice that the $Sp$ dynamics lifts the origin while the $\Psi$ e.o.m.
conflict with that. In other words, along the antibaryon $Sp$ becomes
stronger, the origin in $m_{ab}$ is lifted, but this conflicts with the
e.o.m. constraint that cubic invariants like $A_1\bF_1^j\bF_1^i$ be
zero. In the above equation we have neglected multiplicative corrections
of order $\psi/\bar \phi\ll 1$, this is because we can always rotate our
fields in such a way that only one antibaryon is nonvanishing. A similar
discussion holds for antibaryons which overlap with $\bar P_{1,2}$.
Notice indeed that the K\"{a}hler metric for $\bar \phi$ is
asymptotically flat. Then it is manifest that the above superpotential
does not give asymptotically vanishing vacuum energy for $\bar \phi\to
\infty$. For instance, by imposing $F$-flatness for the mesons we get
the following superpotential for the antibaryons
\eqn\singlets {W\sim \bar \phi^{1+p/(m+1)}\left (Pf \Psi
\right )^{1/(m+1)}.}
According to this equation, at large $\bar \phi$, the fields $\Psi$ are
driven away from the origin. This is because $Pf \Psi$ is the product of
$m+1-p<m+1$ fields whose K\"{a}hler metric is non singular at $\Psi=0$.
However as $\Psi$ is driven away from the origin, the antibaryon
$\bar\phi$ will be pushed towards smaller values by
$|\partial_{\bar\phi} W|^2$. Then we expect that $\bar \phi$ and $\Psi$
will end up being comparable, {\it i.e.} an antibaryon giving mass to
all flavors ends up being excited. In this situation the above analysis
has to be repeated at $p=m+1$, for which there are no light $Sp$
flavors. Then eq. \quantumeff\ is just replaced by $W \sim \bar
\phi^{2+1/(m+1)}$ which pushes $\bar\phi$  back to the origin. We
conclude that the antibaryonic flat directions are lifted by the $Sp$
quantum effects.

\listrefs
\bye